\pdfoutput=1
\documentclass[aoas]{imsart}
\usepackage{amsmath, graphicx, layout, verbatim, enumerate}
\usepackage{enumitem}
\usepackage{lscape}
\usepackage{fancyhdr}
\usepackage{float}
\usepackage{natbib}
\usepackage{verbatim}
\usepackage{subfig}
\usepackage{amsthm}
\usepackage{amsfonts}
\usepackage{amssymb}

\usepackage{algorithm}
\usepackage{algpseudocode}

\startlocaldefs
\endlocaldefs

\newcommand{\red}[1]{\textbf{\color{red} ***ATTN*** #1}}
\renewcommand{\vec}[1]{\mathbf{#1}}

\def\I{{\mathbb I}}
\def\P{{\mathbb P}}
\def\E{{\mathbb E}}
\def\V{{\mathbb V}}

\newtheorem{thm}{Theorem}

\newtheorem{remark}[thm]{Remark}

\theoremstyle{definition}

\renewcommand{\comment}[1]{}

\def\x{{\vec{x}}}
\def\X{{\vec{X}}}

\begin{document}

\begin{frontmatter}

\title{Photo-$z$ Estimation: An Example of Nonparametric Conditional Density Estimation under Selection Bias} 
\runtitle{Nonparametric Density Estimation under Selection Bias}


 \author{\fnms{Rafael} \snm{Izbicki}$^1$,\corref{}\ead[label=e1]{rizbicki@ufscar.br} 
 \fnms{Ann} \snm{B. Lee}$^2$, \ead[label=e2]{annlee@stat.cmu.edu}
 \and 
 \fnms{Peter} \snm{E. Freeman}$^2$\ead[label=e3]{pfreeman@cmu.edu}
 }
 \address{Rafael Izbicki \\ Departamento de Estat\'istica \\ Rodovia Washington Lu\'is, km 235 - SP-310 \\
 S\~ao Carlos - S\~ao Paulo - Brazil \\  \printead{e1}}
\affiliation{$^1$Department of Statistics, Federal University of S\~ao Carlos, Brazil\\
$^2$Department of Statistics, Carnegie Mellon University, USA}


\address{Ann B. Lee and Peter Freeman \\ 5000 Forbes Avenue \\
Department of Statistics, Baker Hall \\
Pittsburgh, PA 15213\\
\printead{e2} \\
\printead{e3}}

\runauthor{R. Izbicki et al.}

\begin{abstract}

Redshift is a key quantity for 
  inferring cosmological model parameters. In photometric redshift estimation, cosmologists use the coarse data collected from the vast majority of galaxies to predict the redshift of individual galaxies. To properly quantify the uncertainty in the predictions, however, one needs to go beyond standard regression and instead estimate the {\em full conditional density} $f(z|\x)$ of a galaxy's redshift $z$ given its photometric covariates $\x$.
   The problem is further complicated by {\em selection bias}: usually only the rarest and brightest galaxies have known redshifts, and these galaxies have characteristics and measured covariates that do not necessarily match those of more numerous and dimmer galaxies of unknown redshift. Unfortunately, there is not much research on how to best estimate complex multivariate densities in such settings. 
  
   Here we describe a general framework for properly constructing and assessing nonparametric conditional density estimators under selection bias, and for combining two or more estimators for optimal performance. We propose new improved photo-z estimators and illustrate our methods on data from the Sloan Data Sky Survey and an application to galaxy-galaxy lensing. Although our main application is photo-$z$ estimation, our methods are relevant to any high-dimensional regression setting with complicated asymmetric and multimodal distributions in the response variable.

\end{abstract}

\end{frontmatter}

\section{Introduction}
\label{sec-intro}

Technological advances over the last two
decades have ushered in the era of ``precision cosmology," with the 
construction of catalogs that contain data on upwards of 10$^8$ galaxies 
(e.g., \citealt{aihara2011eighth}).  Cosmologists use these data to
place progressively tighter
constraints on the parameters of the $\Lambda$CDM model, the
leading model explaining the structure and evolution of the Universe
(see, e.g., \citealt{springel2006large}).  \comment{This data flood will only 
intensify in the next two decades: for instance, the Large Synoptic Survey
Telescope is expected to collect data for up to 10$^{10}$ galaxies during its
ten years of observations that will begin c.~2020 (\citealt{ivezic2008large}).}

To estimate distances to astronomical sources, i.e., to
place them in time relative to the Big Bang,
cosmologists need to determine a galaxy's {\em redshift}, the increase in the wavelength of a traveling photon due to the expansion of the Universe. One can accurately estimate redshift via {\em spectroscopy}, but because of cost and time considerations, 
   more than 99 percent of today's galaxy observations are instead from {\em photometry}, a fast but low-resolution measuring technique 
  where a few broad-band filters coarsely record the radiation from an astronomical object.  
%
%

\begin{figure}[H]
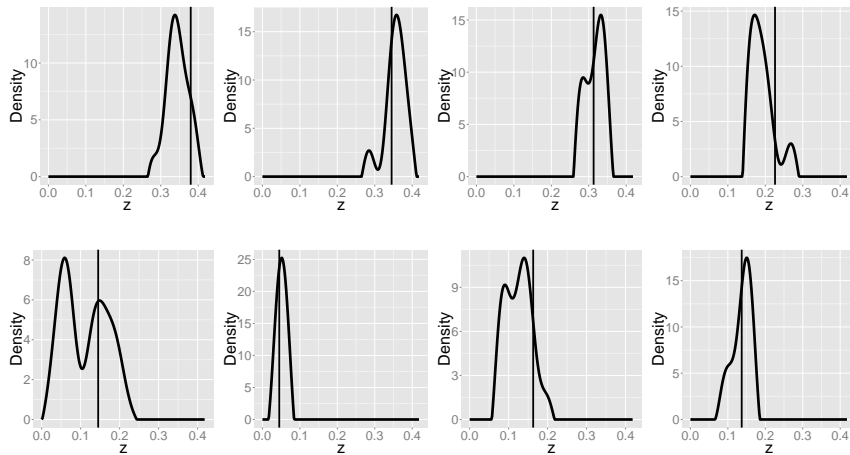

  \centering
\subfloat{\includegraphics[page=10,scale=0.16]{photoZSheldonLargerNScaleFinalForMAPIndividualThesis}}
\subfloat{\includegraphics[page=21,scale=0.16]{photoZSheldonLargerNScaleFinalForMAPIndividualThesis}}
\subfloat{\includegraphics[page=22,scale=0.16]{photoZSheldonLargerNScaleFinalForMAPIndividualThesis}}
\subfloat{\includegraphics[page=4,scale=0.16]{photoZSheldonLargerNScaleFinalForMAPIndividualThesis}} \\
\subfloat{\includegraphics[page=5,scale=0.16]{photoZSheldonLargerNScaleFinalForMAPIndividualThesis}}
\subfloat{\includegraphics[page=17,scale=0.16]{photoZSheldonLargerNScaleFinalForMAPIndividualThesis}}
\subfloat{\includegraphics[page=14,scale=0.16]{photoZSheldonLargerNScaleFinalForMAPIndividualThesis}}
\subfloat{\includegraphics[page=28,scale=0.16]{photoZSheldonLargerNScaleFinalForMAPIndividualThesis}}
  \caption{Photometric estimates of $f(z|\x)$ for eight randomly chosen galaxies from the Sloan Digital Sky Survey (SDSS). Here we use the \textit{Series} estimator from Sec.~\ref{sec-CDEunderCS}. Many of these densities are highly asymmetric and multimodal. The vertical lines indicate the true redshift, determined via high-resolution spectroscopy.}
  \label{fig::photoZ}
\end{figure}

The goal of {\em photometric redshift estimation} (or photo-$z$ estimation) is to estimate 
 a galaxy's redshift $z$ given its photometric covariates $\vec{x}$. 
 Traditionally, redshift estimation has been viewed as a regression problem where one seeks 
  the conditional mean $\E[z|\vec{x}]$ or the most probable redshift of a galaxy.  Recent work, however, shows the importance of  estimating the {\em full conditional density} $f(z|\vec{x})$ \citep{Ball}.
   In photometry,
%
%
$f(z|\vec{x})$ is often asymmetric, multi-modal, with errors that are  heteroscedastic. Fig.~\ref{fig::photoZ} shows density estimates for eight  randomly chosen galaxies from the  Sloan Digital Sky Survey (SDSS): these distributions are complicated non-Gaussian distributions which cannot easily be summarized by, for example, means and variances.
 By working with a density estimate $\widehat{f}(z|\vec{x})$, instead of a point estimate of $z$,  one can dramatically reduce systematic errors in downstream  analysis, i.e., when estimating  functions $g(z)$ of an unknown redshift $z$; see, for example, \citealt{mandelbaum,Wittman,Sheldon}. 
Although optimizing the function $g$ would be optimal for the problem of choice, often there is no clear $g$ ahead of time or there exist many different functions $g$ for the same application;  
 for example, for the galaxy-galaxy lensing 
problem in Sec.~\ref{sec-lensing}, each of  $\approx 500,000$ lenses has a different calibration function $g$.
Hence, a common practice in astronomy is to build catalogs of photo-$z$ density estimates (for galaxies in a survey), which can then be used to address a range of different inference problems in astronomy and cosmology. Many of these estimates, however, include handpicked tuning parameters and, as we later show in Sec.~\ref{sec-lensing} (Fig. \ref{fig::estimatesL}), estimates that yield good results for a particular application or function of $z$ do not necessarily predict the redshift $z$ per se well. This brings up the question of how to properly construct and assess conditional density estimators, in general, and photo-$z$ estimators, in particular.

  
  On the methodological side, photo-$z$ estimation presents challenges at the boundaries of current statistical research. There exist a wide range of sophisticated techniques for high-dimensional inference but few attempt to estimate full conditional densities or ratios of high-dimensional densities (see, e.g. \citealt{IzbickiLeeSchafer,IzbickiLee_CDE} and references within). Even less is known about how to estimate conditional densities {\em under selection bias}. Statistics and machine learning procedures typically assume that training and test data have similar distributions but the two distributions can be very different for sky surveys that mix spectroscopy and photometry, and e.g. remote sensing applications where different sensors may malfunction at different rates or collect data at different rates  (see \citealt{Moreno_Torres}  for data set shift in {\em classification}).
  
   In the astronomy literature, existing photo-$z$ algorithms roughly fall into two categories.
 In the first category, {\em template fitting} (e.g., \citealt{Fernandez2}) estimates $f(z|\vec{x})$   by directly comparing observed data with a suite of idealized photometric data sets for different types of galaxies 
at different redshifts.  Our interest lies in the
second category, {\em empirical redshift estimation}, in which
 one builds an estimator of $f(z|\vec{x})$ using a training set of galaxies with spectroscopically confirmed redshifts 
 (see, e.g., \citealt{Ball,Zheng,kind2013tpz}).  
 In photo-$z$ estimation, however, high-resolution spectroscopy is extremely time-consuming. For example, one would already need $\sim$2.5 years of dedicated telescope time to estimate the spectroscopic redshifts for the 500{,}000-galaxy photometric SDSS sample in our paper; the problem only gets worse with deeper and larger surveys. As a result, 
 usually only the rarest and brightest galaxies are spectroscopically observed, and these galaxies have characteristics and measured covariates that do not necessarily match those of more numerous dimmer galaxies  (see, e.g., \citealt{Oyaizu,Ball}). 

The goal of this paper is to develop improved nonparametric methods for conditional density estimation that explicitly deal with photo-$z$ estimation and selection bias:
\begin{enumerate}
\item On the methods side, we present a general framework for supervised learning that
 accounts for differences in training and test data, and that unlike standard nonparametric regression and classification can handle highly asymmetric and multimodal distributions. 
  We design appropriate loss functions for a multivariate setting with selection bias (Eqs. \ref{eq-lossBetaEst} and \ref{eq-lossCDEEstimate}), and we show how to estimate these functions using labeled and unlabeled data. Our set-up provides a principled method for choosing tuning parameters, for selecting covariates, and for comparing and combining (Sec.~\ref{sec-combine}) different conditional density estimators for optimal performance. The final density estimates lead to more accurate {\em predictive intervals}  (see, for example, Fig. \ref{fig::estimatesL}) for new observations.
  Estimating $f(z|\x)$ is also a simple way of performing {\em nonparametric quantile regression}  of many quantiles simultaneously \citep{Koenker2005}.

 \item In the context of photo-$z$ prediction, we propose more accurate algorithms for estimating photo-$z$ probability distributions in a setting where primarily nearby and bright galaxies have known redshifts.
  Using SDSS as well as simulated data, we analyze and compare different methods for estimating conditional densities and so-called importance weights (which are used to correct for selection bias). 
We introduce two new conditional density estimators, \emph{Kernel nearest neighbors} (Sec.~\ref{sec-ker-NN}) and \emph{Series} (Sec.~\ref{sec-CDEunderCS}), that both have better performance than the photo-z prediction method by \citet{Cunha1}, 
which represents the state-of-the-art for empirical photo-$z$ estimators under selection bias.  We also
  present different diagnostic tests for evaluating the goodness-of-fit of estimated densities (Appendix~\ref{app:diag}). \end{enumerate} 

\comment{In particular, we propose two alternative non-parametric estimators 
to the conditional density estimator of \citet{Cunha1}. 
We compare our new estimators to that of \citet{Cunha1} using SDSS data and
demonstrate that they yield improved results.
Moreover, we introduce a principled method of combining 
two or more conditional density estimators for optimal performance under 
different sampling schemes.  We compare various methods for estimating 
importance weights -- an important quantity used to correct methods so that they work under selection bias -- and describe how to test the goodness-of-fit of the 
final density estimates.}

The organization of the paper is as follows: Sec.~\ref{sec-data} describes our data. In Sec.~\ref{sec-CS}, we introduce the statistical problem and the idea of importance weights. Sec.~\ref{sec::appliWeights} compares  different schemes for estimating these weights. Then, in Sec.~\ref{sec-densityUnderCS}, we shift our focus to the problem of estimating conditional densities under selection bias, and the issue of proper model selection and assessment. Finally, in Sec.~\ref{sec-lensing}, we offer some new insights on the galaxy lensing-lensing example from \citet{Sheldon}. We conclude our work with Sec.~\ref{sec-conclusions}.  

\section{Data}
\label{sec-data}

There are two main types of astronomical data:  \emph{spectroscopic} data, where both the covariates $\x$ and the redshift $z$ (the label) can be measured with negligible error, and (ii)  \emph{photometric} data, where only the covariates $\x$ are known and there is no precise measurement of the redshift. In our study, we use photometric and spectroscopic data from the Sloan Digital Sky Survey (SDSS; \citealt{york00}), as well as SDSS-based simulated data with known levels of selection bias.
 
\comment{Since 1998, th has
collected data on over 200 million galaxies that are spread across one-quarter
of the sky. There are two basic types of data that SDSS collects:
\emph{spectroscopic}
data, from which a galaxy's redshift $z$ (an observable proxy for its distance 
from the Earth) may be directly measured with negligible error; and
\emph{photometric} data, which does not allow for such precise measurement
of redshift. Since spectroscopic data for a given galaxy takes far longer to 
collect than photometric data, the number of galaxies observed spectroscopically
is relatively small ($\sim$1 million). The goal of photometric redshift
estimation, or photo-$z$ estimation, is to train an algorithm using the
photometric data of galaxies with known spectroscopic redshifts.
We more fully describe both types of data below.}
  
\subsection{SDSS Photometric Data} \label{sec-sdss}
 Since 1998,  SDSS has  
  collected data on over
200 million galaxies that are spread across one-quarter of the sky. The vast
majority ($\gtrsim$99\%) of these galaxies are only photometrically observed.

In photometry, different filters are sequentially placed into
a telescope's light path. Each filter only allows passage of photons in a particular wavelength band. SDSS measures photon 
fluxes, or equivalently \emph{magnitudes} (the logarithm of fluxes), in 
five bands, denoted $u$, $g$, $r$, $i$, and $z$, in the wavelength range 
$3.5 \times 10^{-7}$ meters to $9 \times 10^{-7}$ meters (i.e., from UV
light through the optical regime to infrared light). 
Magnitude estimates are algorithm-dependent, in that they depend on how one defines a boundary around a galaxy and how one sums the light within that boundary. SDSS catalogs include estimates from many different boundary-definition algorithms or {\em magnitude systems}; in this work, we follow \citet{Sheldon}
and use {\tt model} and {\tt cmodel} magnitudes.  We also work with \emph{colors}, or 
  differences of magnitudes 
  in adjacent wavelength bands. 
  
  Our final SDSS photometric data set contains 10 covariates 
(four colors and the associated $r$-band magnitude from each algorithm) for  
538,974 galaxies in an $\approx$72 square-degree sky patch.\footnote{
Celestial longitude, or right ascension (RA) $\in$ [168$^{\circ}$,192$^{\circ}$]
and celestial latitude, or declination ($\delta$) $\in$
[$-1.5^{\circ}$,$1.5^{\circ}$].}
 These galaxies are extracted from SDSS Data Release 8 (DR8; \citealt{aihara2011eighth}) and
filtered according to \citet{Sheldon}. Each covariate is normalized to have mean 0 and standard deviation 1.

\subsection{SDSS Spectroscopic Data} \label{sec-sdss-labeled}
Of the over 200 million 
galaxies in SDSS, some one million have been the subject of follow-up 
{\em spectroscopic} observations. In spectroscopy, a light-dispersing grating or
prism is placed into a telescope's light path. The photon changes its path with an angle that is proportional to its wavelength. Thanks to high-resolution mapping of the dispersed light, one can finely resolve narrow spectral features (or lines) that are smeared out in
photometry. These lines, which are caused by transitions of electrons between atomic energy levels,
occur at known wavelengths $\lambda$ but are observed at wavelengths
$(1+z)\lambda$, where $z$ is the galaxy's redshift.\footnote{In astronomy, the notation 
  $z$ is used to denote both
redshift and a particular photometric band. For the remainder of this work, $z$ will always represent redshift.}
One can use the wavelength ratios of two or more observed lines to infer 
which transitions they represent; once that information is known, redshift
estimation is trivial. 
  The precision in the estimates is typically  
${\Delta z}/z \sim 10^{-6}$, i.e., for spectroscopic redshifts, we can safely ignore the measurement error.

In this paper, we use the same   data  
(i.e., colors and $r$-band magnitudes, collectively denoted $\x$, and 
redshifts $z$) as in \citet{Sheldon}. This data set 
includes 435,875 galaxies; the vast majority are taken from SDSS DR8 but
 some fainter galaxies have been added so that the spectroscopic covariates cover 
the same space as the photometric covariates, albeit with a different distribution
  (E. Sheldon, private communication).

\begin{figure}[H]
  \centering
\subfloat{\includegraphics[page=1,scale=0.33]{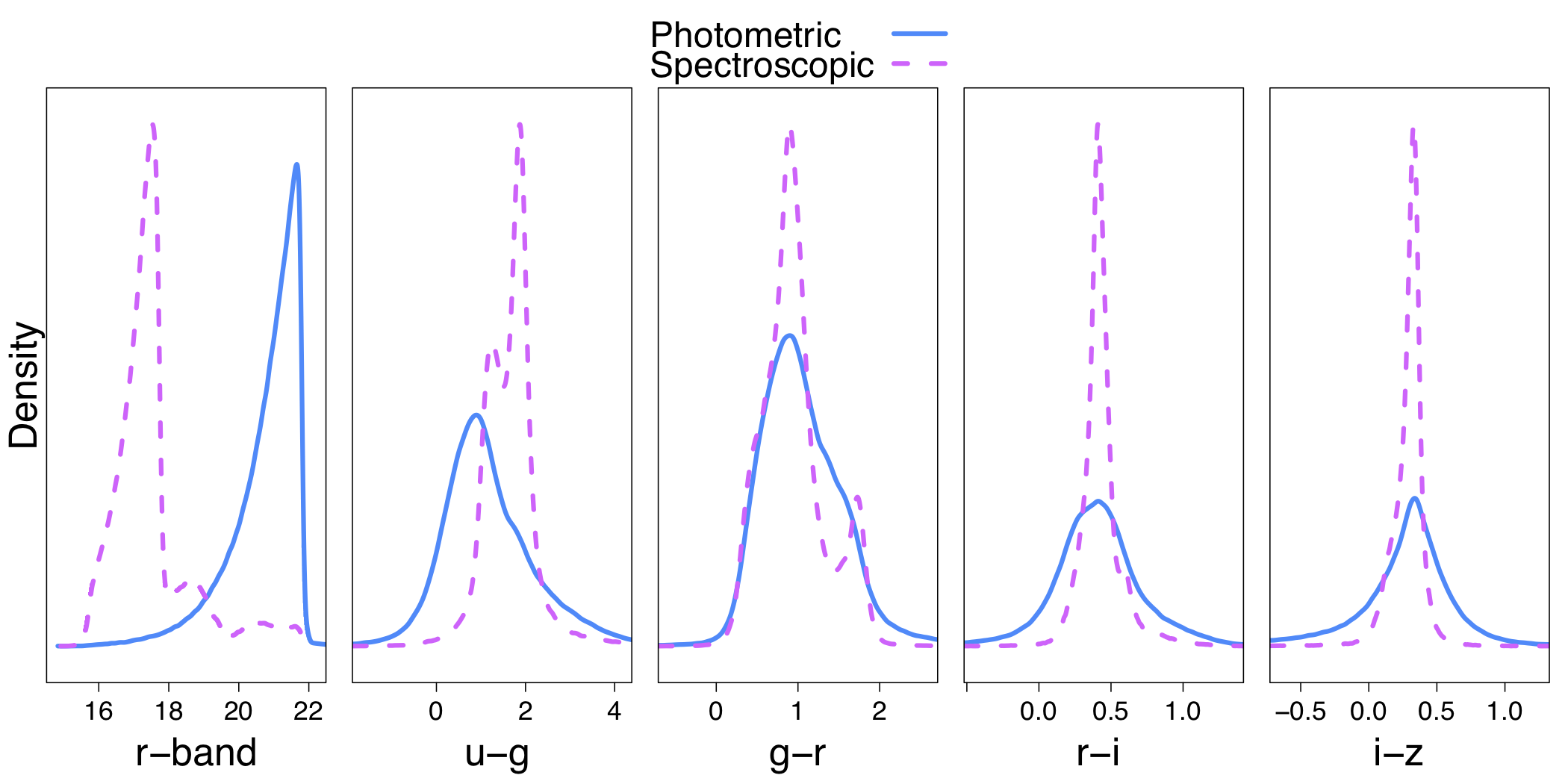}}\hspace{-2mm}
  \caption{\small Distribution of $r$-band {\tt model} magnitude and colors (i.e., differences of {\tt model} magnitude values in adjacent photometric bins) for photometric versus spectroscopic SDSS data. Many galaxies in the spectroscopic sample are brighter (i.e. they have a lower  $r$-band magnitude as indicated in the left panel) than the galaxies in the photometric sample.}
  \label{selectionBiasReal}
\end{figure}

Fig.~\ref{selectionBiasReal} shows the distributions of the spectroscopic and photometric samples. The left panel shows that there is a clear selection bias in the study where the galaxies in the spectroscopic sample tend to be brighter (i.e. they tend to have a lower $r$-band magnitude) than the galaxies in the photometric sample. 

\subsection{Simulated Samples with Known Levels of Selection Bias}  
\begin{figure}[H]
\centering\subfloat{\includegraphics[page=1,scale=0.33]{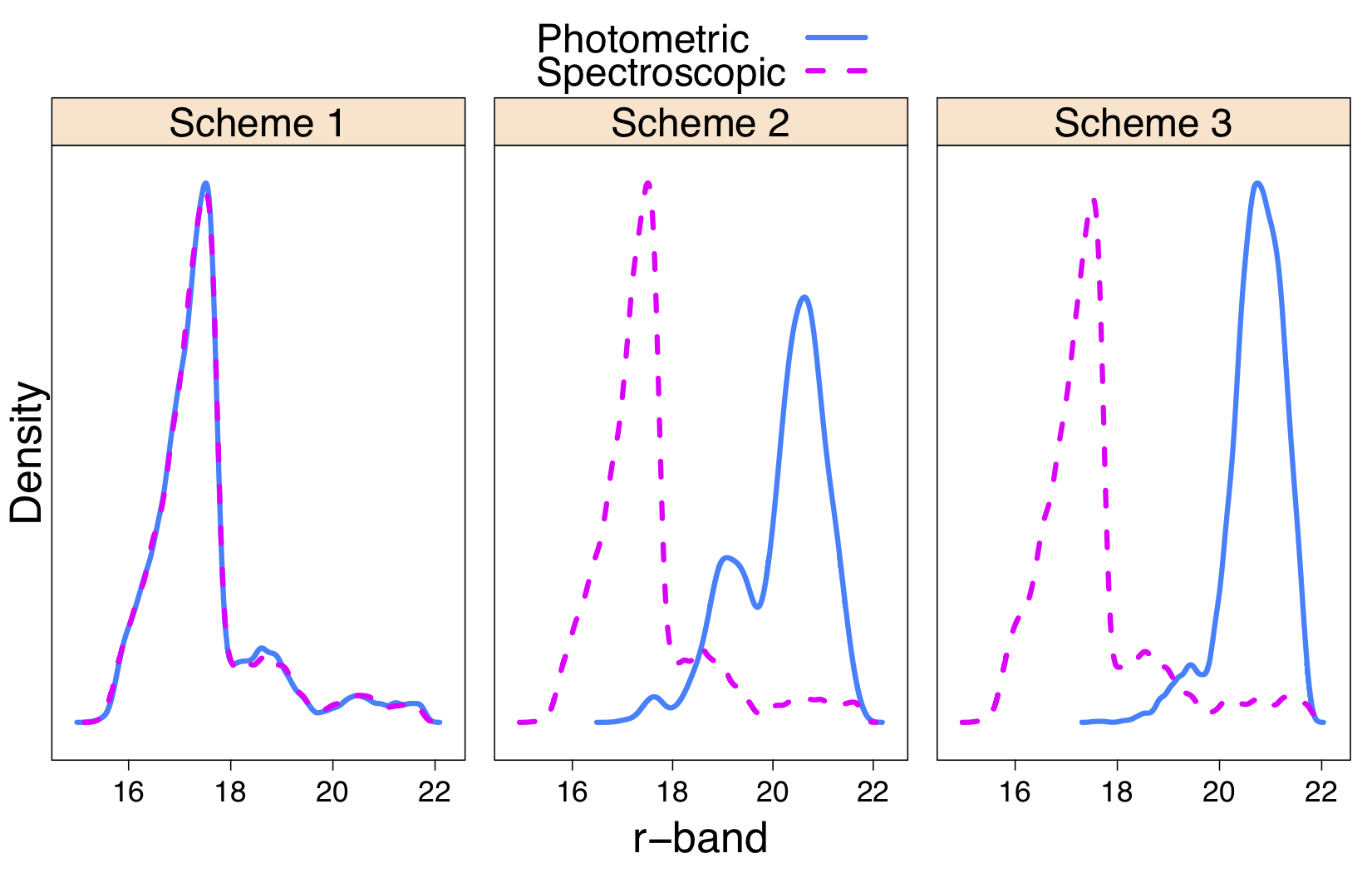}}
 \caption{\small Distribution of the $r$-band {\tt model} magnitudes for 
 three different schemes of resampling the SDSS spectroscopic sample. The simulated data sets have (known) levels of selection bias realistic for different astronomical surveys.}
 \label{selectionBias}
\end{figure}

In addition to the SDSS data, 
we construct photometric samples with {\em known} levels of selection bias relative to the SDSS spectroscopic data. We use the following rejection sampling algorithm:   Let $\vec{x}$ be a data point in the spectroscopic sample with $r$ {\tt model} magnitude $x^r$ (scaled to be between $0$ and $1$). 
A large $r$ {\tt model} magnitude corresponds to a faint galaxy. 
 Let $S$ be a binary selection variable that decides whether a galaxy in the spectroscopic sample is included in the photometric sample ($S=1$) or not ($S=0$). We assume that  the probability $P(S=1|\vec{x})$ depends on $\vec{x}$ through $x^{(r)}$ only; i.e., that $P(S=1|\vec{x})=P(S=1|x^{(r)})$.
 To create  levels of selection bias
 realistic for different astronomical surveys, we resample the SDSS spectroscopic sample according to\\
 \indent Scheme 1: $P(S=1|x^{(r)}) = f_{\mbox{B(1,1)}}(x^{(r)}) \equiv 1$,\\
 \indent Scheme 2: $P(S=1|x^{(r)}) = f_{\mbox{B(13,4)}}(x^{(r)}) $,\\
 \indent Scheme 3: $P(S=1|x^{(r)})= f_{\mbox{B(18,4)}}(x^{(r)})$,\\
 where $f_{\mbox{B(i,j)}}$ denotes the density of a beta random variable
 with parameters $(i,j)$.
Fig.~\ref{selectionBias} shows the resulting $r$-band distributions. 
 Scheme 1 involves no selection bias: the
spectroscopic and photometric data have $r$-band magnitudes with the same 
distribution.  At the other extreme is Scheme 3 with strong selection bias: most photometrically observed galaxies are significantly fainter (shifted toward large $r$-band magnitude) than the galaxies in the spectroscopic data set. 

\section{Selection Bias, Covariate Shift and Importance Weights} \label{sec-CS}
 
 A standard assumption in statistics and machine learning is that labeled and unlabeled data have similar distributions but, as Fig.~\ref{selectionBiasReal} (left panel) shows, the two distributions can be very different for sky surveys that mix spectroscopy and photometry: Brighter galaxies (or galaxies with a lower $r$-band magnitude) are more likely to be selected for follow-up spectroscopic observation. This fact motivates the methods and algorithms presented in this paper. 
 Below, we fix our notation and describe the main ideas behind importance weights as a way of correcting for sample selection bias. 
 
\subsection{Problem Statement and Notation}
\label{sec-pform}
Our data are covariates $\vec{x} \in \mathbb{R}^d$ (photometric colors and magnitudes) and labels $z \in \mathbb{R}$ (the redshift). Without loss of generality, we rescale the original redshift values so that the response $z\in[0,1]$.
 
Suppose we have access to an i.i.d unlabeled sample  $\vec{x}^U_1,\ldots,\vec{x}^U_{n_U}$ with only photometric data, and an i.i.d. labeled sample  $(\vec{x}^L_1,z^L_1),\ldots,(\vec{x}^L_{n_L},z^L_{n_L})$ from a potentially different distribution, where the labels are from follow-up spectroscopic studies. Because the high cost of spectroscopy in sky surveys, in terms of telescopic resources, $n_L \ll n_U$ and the distributions of the labeled and unlabeled samples will {\em not} be the same.
Our goal in this paper is to construct a  photo-$z$ density estimator $\widehat{f}(z|\vec{x})$ that performs well on the {\em unlabeled} photometric data, which roughly dominate 99\% of today's galaxy observations.


To fix our notation, let $\P_{L}$ and $\P_{U}$ denote the distributions on the labeled and 
unlabeled samples, respectively; i.e., let $(\vec{x}^L,z^L)\sim \P_L$ and 
$(\vec{x}^U,z^U)\sim \P_U$, where $z^U$ are missing data labels. We say that there is a 
\emph{data set shift} if $\P_{L}\neq \P_{U}$.  To understand how this affects learning algorithms, one needs to make additional assumptions about the relationship between $\P_{L}$ and $\P_{U}$ (e.g., \citealt{Quionero-Candela:2009:DSM:1462129},
\citealt{Gretton}, \citealt{Moreno_Torres}).  For our application,
we assume that the probability that a galaxy is labeled with a spectroscopic
redshift is independent of the response variable $z$ if we condition on the covariates $\vec{x}$ (\citealt{Lima1, Sheldon});
 i.e., 
\begin{align}
\label{eq-CSassumption}
\P(S=1|\vec{x},z)=\P(S=1|\vec{x}) \ ,
\end{align}
where the random variable $S$ equals $1$ if a datum is labeled and $0$ 
otherwise. This type of sample selection bias where Eq.~\ref{eq-CSassumption} holds is sometimes referred to as \emph{missing at random} (MAR;
\citealt{Rubin1976Inference}, \citealt{Moreno_Torres}) bias.\footnote{MAR is to be distinguished from data missing completely at random (MCAR) which occurs when the sampling method is completely independent of $\vec{x}$ and $z$, i.e., 
$\P(S=1|\vec{x},z)=\P(S=1)$, and hence there is no data set shift.} 
MAR bias
 implies \emph{covariate shift}, defined as 
\begin{equation}\label{eq:covariate_shift}
f_{L}(\vec{x}) \neq f_{U}(\vec{x}), \ \  f_L(z|\vec{x})=f_U(z|\vec{x}).
\end{equation}
Under certain conditions on the support of $f_L(\vec{x})$
and $f_U(\vec{x})$, covariate shift also implies MAR bias
(\citealt{Moreno_Torres}). 
Below we use the terms interchangeably 
to refer to the assumption in Eq.~\ref{eq-CSassumption}. 

\begin{figure}
    \includegraphics[page=1,scale=0.29]{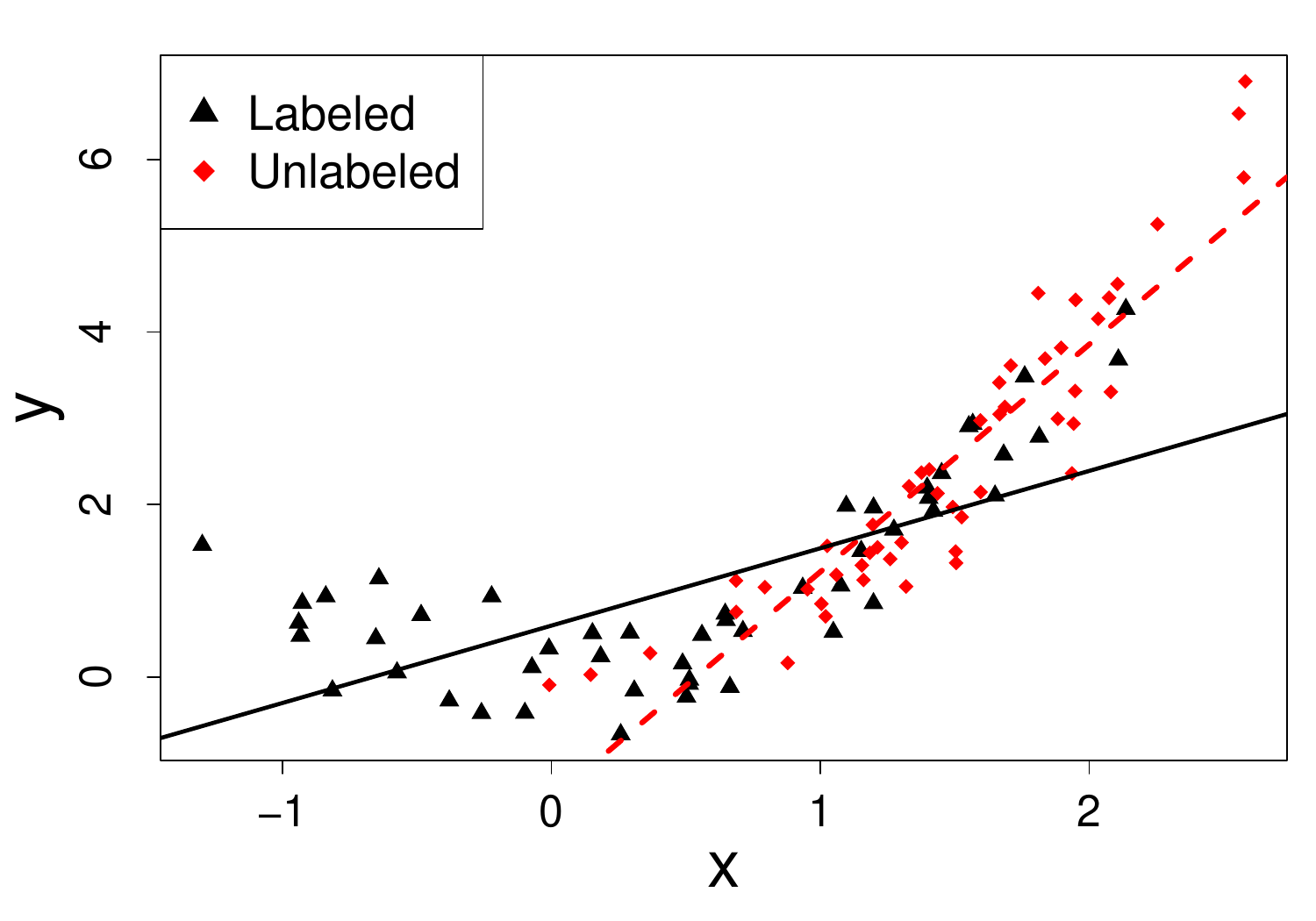}
    \caption{\small  Toy example for covariate shift in linear regression: The solid line is reasonable for predicting new data from the labeled sample, but it is far from  optimal for the unlabeled sample.}
    \label{fig::csExample}
\end{figure}

At first glance, it may seem that 
MAR bias 
 would not pose a problem for density estimation: Because $f(z|\vec{x})$ is the same for both labeled and unlabeled samples (Eq.~\ref{eq:covariate_shift}), one might conclude 
that a good estimator of $f(z|\vec{x})$ based on labeled data would also have good performance for unlabeled data. This is in general not true. 
Often $f(z|\vec{x})$ is well estimated only in regions where there is plenty of labeled data and these regions may not coincide with regions where there is plenty of unlabeled (target) data; see 
 Figure~\ref{fig::csExample} for a toy example of an equivalent regression problem with covariate shift. 
  From a statistical
perspective, this problem arises because 
the loss function used for 
estimating $f(z|\vec{x})$ (or analogously $\mathbb{E} [z|\vec{x}]$ in regression) 
depends on the {\em marginal} distribution of
$\vec{x}$. Hence, {\em an estimator that performs well 
with respect to $f_{L}(\vec{x})$ may not perform well
with respect to $f_{U}(\vec{x})$}.

One solution to this mismatch problem is to reweight the labeled data so that their distribution $\P_{L}$ -- after a reweighing with the so-called  {\em importance weights} $\beta(\vec{x}):=f_{U}(\vec{x})/f_{L}(\vec{x})$ -- matches the distribution  $\P_{U}$ of the unlabeled data. We can then compute expectations with respect to the target distribution $\P_{U}$ using the distribution  $\P_{L}$ of the labeled training data. In particular, if $l(\widehat f, f)$ is the loss function for the estimated conditional density $\widehat f$, we use that 
 \begin{equation} \mathbb{E}_{(\X,Z) \sim \P_{U}} \left[l(\widehat f(Z|\X), f(Z|\X))\right] = \mathbb{E}_{(\X,Z) \sim \P_L} \left[ l(\widehat f(Z|\X), f(Z|\X))\beta(\X) \right]. \label{eq:importance_weighting}\end{equation}
\comment{\red{Ann: Should we change the notation/choice of words in Sec 5 and refer to the expected loss in Eq. (8) as the "risk" R and the modified risk as  J and so on? Maybe it's confusing to use the term loss and big L for non-random quantities. We also use L to denote labeled data but that's a minor problem.}
\red{Rafael responds: I think typically people think of the risk as the expectation wrt the training sample well, so I guess that
could be confusing too. As the referees didn't mention that, I suggest we leave it as it is for now.}}

 The last expression of Eq.~\ref{eq:importance_weighting} involves unknown weights $\beta(\x)$. 
Open questions are:
\begin{enumerate}
\item how to best estimate the importance weights $\beta(\vec{x}):=f_{U}(\vec{x})/f_{L}(\vec{x})$, where $\x$ is our multivariate data,
 and
\item how to design an estimator of the full conditional density $f(z|\vec{x})$ that performs well on the target data; i.e., that has a small risk according to Eq.~\ref{eq:importance_weighting}. 
\end{enumerate}
In this work, we propose a greedy two-step approach to conditional density estimation under selection bias where one first selects the best model for estimating $\beta(\vec{x})$ (see Sec.~\ref{sec::appliWeights}), and then uses these estimates to search for the best model for estimating $f(z|\vec{x})$ under covariate shift (see Sec.~\ref{sec-densityUnderCS}). Note that a necessary condition for computing importance weights is that $\P_{L}$ {\em dominates} $\P_{U}$, i.e.,  $\P_{L}(\vec{x})\gg\P_{U}(\vec{x})$; hence, for this study, we have selected photometric data whose covariates lie within the domain of the spectroscopic covariates (see Sec.~\ref{sec-sdss}).

\section{Estimating Importance Weights}
\label{sec::appliWeights}

\comment{In this section, we describe common estimators of importance weights,
$\beta(\vec{x})$. We then propose a
new weighted loss function, which we use to demonstrate the relative
superiority of the nearest neighbor- and spectral series-based estimators 
of \cite{Lima1} and \cite{IzbickiLeeSchafer}, respectively.}


A naive method for computing $\beta(\vec{x})$ is to separately
estimate $f_U$ and $f_L$ and to then take their ratio, but this approach can enhance errors in the individual density estimates, particularly in regions where $f_L$ is nearly zero \citep{SugiyamaImportance}. There are also several other approaches for estimating density ratios (see Appendix~\ref{app:bias}). 

The goal in this section
is to find out how these estimators perform in practice on our data. To assess the estimators, we use the weighted loss function by \cite{IzbickiLeeSchafer} defined as 
\begin{align}
\label{eq-lossBeta}
L(\widehat{\beta},\beta):&=\int \left(\widehat{\beta}(\vec{x})-\beta(\vec{x}) \right)^2dP_L(\vec{x}) \notag \\ 
&=\int \widehat{\beta}^2(\vec{x}) dP_L(\vec{x})-2\int \widehat{\beta}(\vec{x}) dP_U(\vec{x})+K ,
\end{align}
where $K$ is a constant that does not depend on $\widehat{\beta}(\vec{x})$. (Appendix~\ref{app:bias} gives 
some intuition behind the choice of this loss function.)
 We divide our data into three parts: a training set used to fit the model, a validation set for model selection and tuning of parameters, and a test set for assessing the final model
\citep[p. 222]{trevor2009elements}. 
For model selection and model assessment, we estimate $L(\widehat{\beta},\beta)$ according to \cite{IzbickiLeeSchafer}:
\begin{align}
\label{eq-lossBetaEst}
\widehat{L}(\widehat{\beta},\beta) =\frac{1}{\widetilde{n}_L}\sum_{k=1}^{\widetilde{n}_L}\widehat{\beta}^2\left(\widetilde{\vec{x}}_k^L\right)-2\frac{1}{\widetilde{n}_U}\sum_{k=1}^{\widetilde{n}_U}\widehat{\beta}\left(\widetilde{\vec{x}}_k^U\right),
\end{align}
 where $\widetilde{\vec{x}}^L_1,\ldots,\widetilde{\vec{x}}^L_{\widetilde{n}_L}$ represent the labeled (validation or test) data, and $\widetilde{\vec{x}}^U_1,\ldots,\widetilde{\vec{x}}^U_{\widetilde{n}_U}$ represent the unlabeled (validation or test) data.

{\em Experiments.}  
We use this loss function to compare six different estimators of $\beta$:
\begin{itemize}
\item \emph{$\beta$-NN}, the nearest neighbor estimator from \citet{Lima1}  (Eq.~\ref{eq-nnWeights} in the Appendix \ref{app:bias}), but 
with the number of nearest neighbors $M$ chosen so as to minimize our estimated loss 
(Eq.~\ref{eq-lossBetaEst}) on the validation data (as in \citealt{kremer2015nearest}); 
\item \emph{$\beta$-NN1}, the nearest neighbor approach with $M = 1$ \citep{Loog}; 
\item \emph{$\beta$-KLIEP} and \emph{$\beta$-uLSIF},
importance weight estimators suggested by \citet{SugiyamaImportance} and 
\citet{Kanamori}, respectively, and implemented with the authors' MATLAB code;\footnote{http://sugiyama-www.cs.titech.ac.jp/~sugi/software}
\item \emph{$\beta$-KuLSIF}, a kernelized version of \emph{$\beta$-uLSIF} \citep{kanamori2012statistical}; and 
\item \emph{$\beta$-Series}, the density ratio estimator from \citet{IzbickiLeeSchafer}.
\end{itemize}
Following \citet{Lima1}, our covariates are the four colors and the $r$-band magnitude in the {\tt model} magnitude system. 

\begin{figure}[H]
\centering
\subfloat{\includegraphics[page=1,scale=0.30]{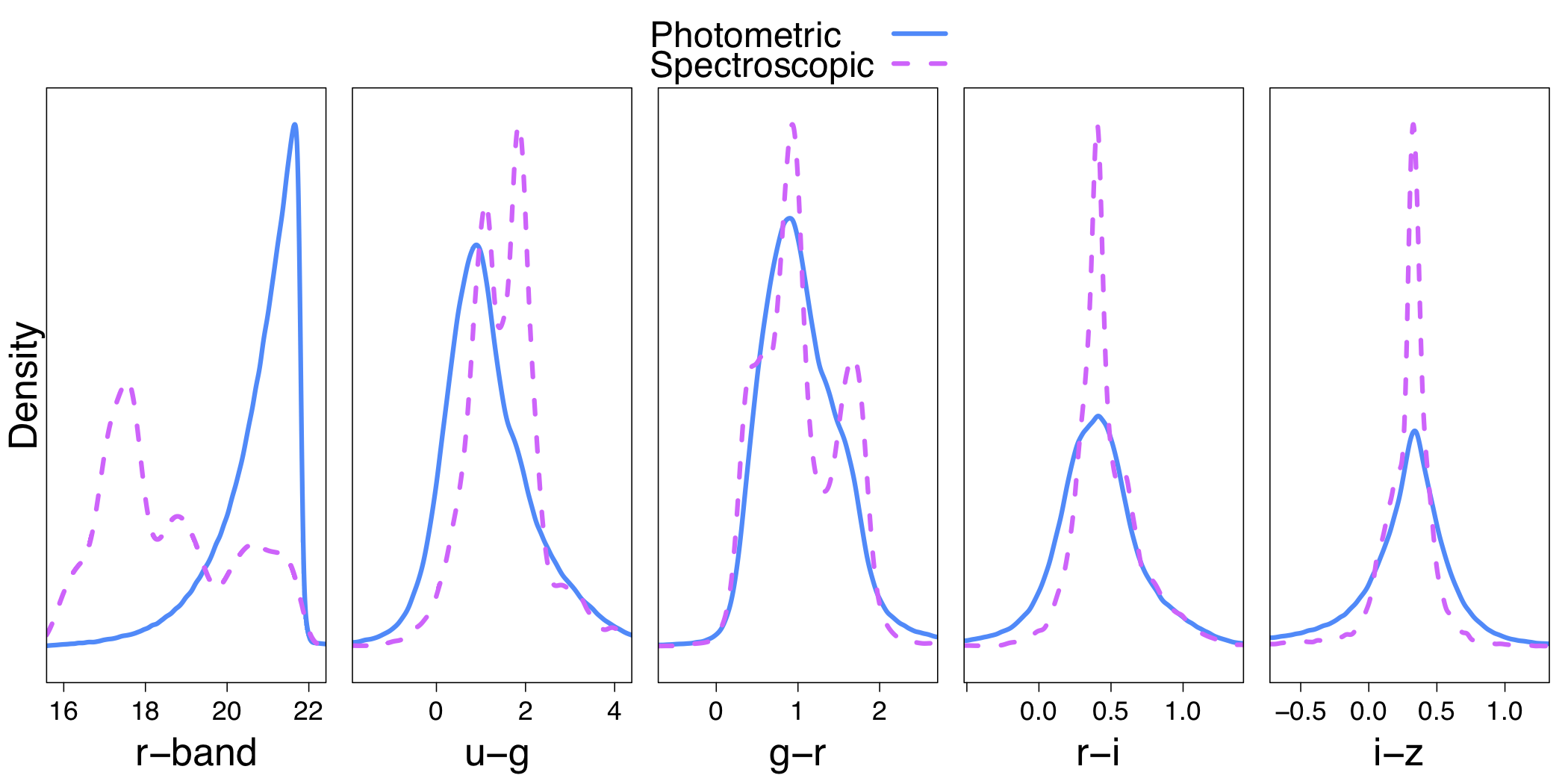}}\hspace{-2mm}
\caption{\small Distribution of the $r$-band {\tt model} magnitude and the 
four colors from {\tt model} magnitude for the SDSS photometric and 
spectroscopic samples \emph{after} we remove and replace labeled examples for which the initial estimates of the importance weights are zero; compare with Fig.~\ref{selectionBiasReal}. By restricting the labeled data to the regions of interest (i.e. to the regions where there are unlabeled data) we increase the effective sample size.}
\label{fig::selectionBiasRealBeta0}
\end{figure}

We study the estimators under the simulated selection bias settings from Section~\ref{sec-sdss} (Schemes 1-3), 
using labeled and unlabeled samples that are each of size 10,000. 
(For each sample, we randomly choose 2,800 galaxies for training, 1,200 for
validation and 6,000 for testing.) We also
apply the estimators to the SDSS data. These data have a large covariate shift (Fig.~\ref{selectionBiasReal}).
where $\widehat{\beta}(\vec{x})=0$ for $\approx$80\% of the labeled examples, 
i.e., the majority of the spectroscopic sample lie in regions of
covariate space where there are no unlabeled data. Having more labeled data may seem harmless but it turns out that if one includes these labeled examples in the training sample, the {\em effective sample size}  (defined in \citealt{Shimodaira2000} and \citealt{Gretton}) will be 
very small, ultimately resulting in poor conditional density estimates 
(see Sec.~\ref{sec:CDE_experiments} and Fig.~\ref{fig::finalestimatesReal}). Essentially, many galaxies have zero weight (i.e. no contribution) in the conditional density estimation. To mitigate this problem, we use a similar data cleaning step as in  \citet{Lima1}: 
First, we construct a $\emph{$\beta$-NN}$ estimator using a preliminary sample of 10,000 spectroscopic and 10,000 photometric (randomly selected) galaxies. Using this estimate, we then create a new spectroscopic sample of size 15,000 which consists of galaxies with weight estimates $\widehat{\beta}(\vec{x})\neq 0$.
The SDSS results in this paper are based on the
{\em new} spectroscopic sample together with 15,000 randomly selected photometric
observations; Fig.~\ref{fig::selectionBiasRealBeta0} shows the distributions of the two samples after the preprocessing step.
In all experiments on SDSS data, we use $3,\!500$ galaxies from each sample (labeled and unlabeled) for training, $1,\!500$ for validation, and the
remainder for testing.

\begin{figure}[H]
\centering
\subfloat[Scheme 1]{\includegraphics[trim={0 1.7cm 0 0},clip,page=1,scale=0.22]{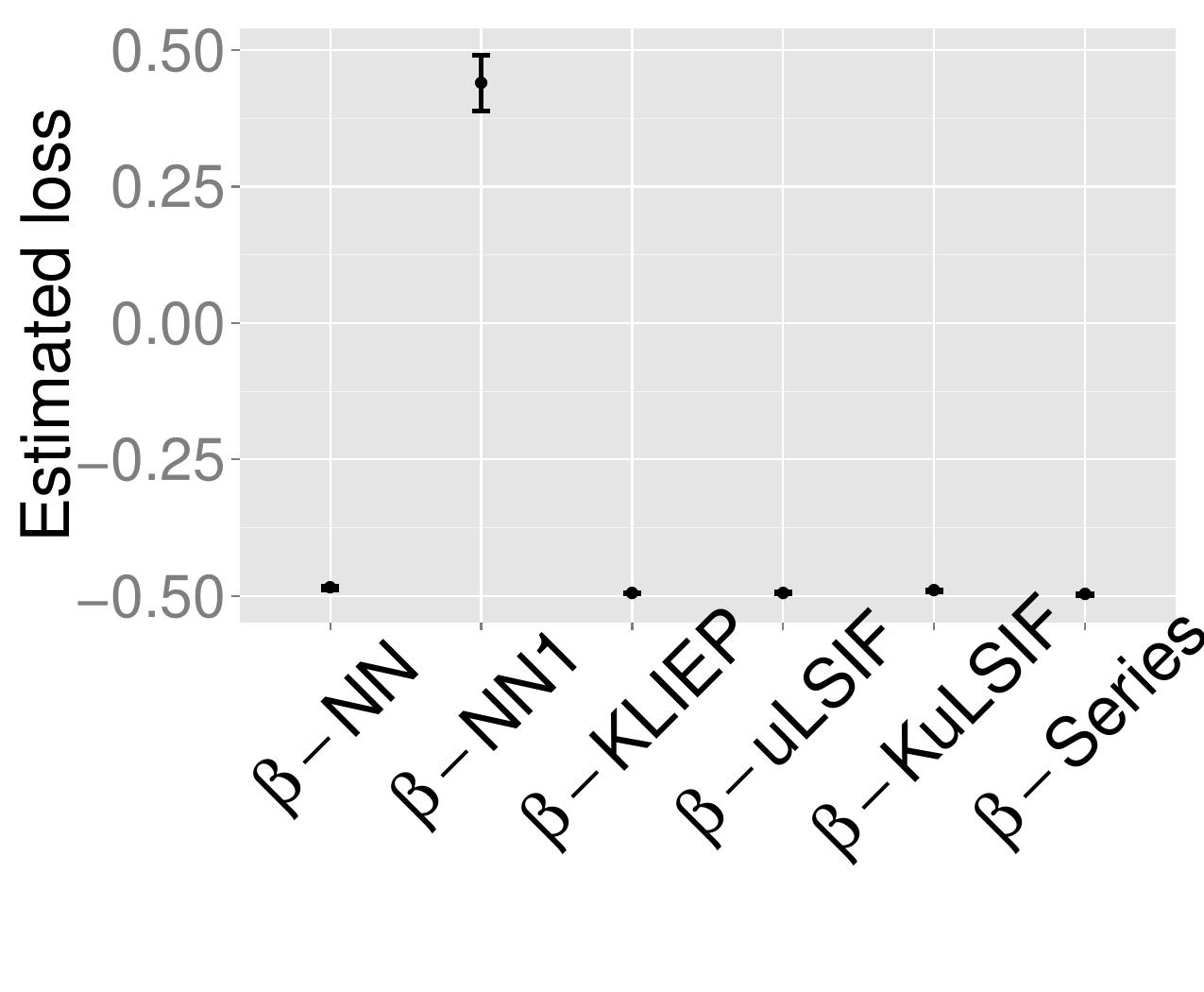}}\hspace{1mm}
\subfloat[Scheme 2]{\includegraphics[trim={0 1.7cm 0 0},clip,page=2,scale=0.22]{GGPLOTsdssBetaEstimationWithErrorsThesis.pdf}}\hspace{-1mm}
\subfloat[Scheme 3]{\includegraphics[trim={0 1.7cm 0 0},clip,page=3 ,scale=0.22]{GGPLOTsdssBetaEstimationWithErrorsThesis.pdf}}\hspace{1mm}
\subfloat[SDSS]{\includegraphics[trim={0 1.7cm 0 0},clip,page=3,scale=0.22]{GGPLOTsdssBetaEstimationReal.pdf}}\hspace{-1mm}
\caption{\small The estimated loss $\widehat{L}(\widehat{\beta},\beta)$ of 
different estimators of $\beta(\vec{x})$ for Schemes 1-3 with, respectively, (a) no, (b) moderate, and (c) large covariate shift. Panel (d) shows $\widehat{L}(\widehat{\beta},\beta)$ for observed
SDSS data. The bars in each panel
correspond to the mean estimated loss $\pm 1$ standard error. These results indicate that \emph{$\beta$-NN} consistently performs the same or better than other competing methods; $\emph{$\beta$-Series}$ is a close second. } 
\label{fig-comparisionWeightsEstimators}
\end{figure}

Fig.~\ref{fig-comparisionWeightsEstimators} compares the different estimators of $\beta$. The nearest-neighbor method is rarely mentioned in the literature on data set shift, but in our experiments we find that  \emph{$\beta$-NN}, where $M$ is chosen by data splitting, consistently performs the same or better than other competing (and more complicated) methods; $\emph{$\beta$-Series}$ is a close second. This is in agreement with recent independent work by \citet{kremer2015nearest}.
We also note that by minimizing Eq.~\ref{eq-lossBetaEst}, we select $M=8$ neighbors for the SDSS dataset, 
a value similar to the value $M=5$ chosen in an ad hoc manner by \citet{Cunha1}.
Henceforth, we will use the \emph{$\beta$-NN} method (defined in Appendix~\ref{app:bias}, Eq.~\ref{eq-nnWeights}) to estimate 
 importance weights. 

{ \em Variable Selection.}  
 One can further improve on these results by variable selection on the full set of 10 covariates. Table~\ref{tab::variableSelecBeta} lists the selected covariates in a forward stepwise model search  with \emph{$\beta$-NN}. For Scheme 1, where there is no selection bias (and where $\beta(\x)$, as a result,  
does not depend on $\x$), the final model includes only one covariate.
In Schemes 2 and 3, the importance weights depend on the {\tt model} $r$-band magnitude, 
and the selected model always includes this covariate.
For the SDSS data, we achieve a loss of $-2.41 \pm 0.08$ with variable selection, which 
is significantly smaller than the value $-2.16 \pm 0.05$ when including all 10 covariates, and the value $-1.97 \pm 0.04$ (see \emph{$\beta$-NN} in Fig.~\ref{fig-comparisionWeightsEstimators}d) when using   
the five covariates from the {\tt model} magnitude system only as in \citet{Sheldon}. 
\comment{\red{added}
We conclude that performing variable selection leads to more accurate estimates of the importance weights. As we discuss
in the next section, this improvement is
key to having good conditional density estimators under covariate shift.} 

\begin{table}[H]
\begin{center}
\caption{\small Selected covariates for estimating importance weights with the $\emph{$\beta$-NN}$ estimator.}
    \begin{tabular}{  l | c |c |c |c |c |c |c |c |c |c }
    \hline
    Data set &  \multicolumn{5}{c|}{{\tt model}}  &  \multicolumn{5}{c}{{\tt cmodel}}  \\ \hline
    & $r$ & $u-g$ & $g-r$ & $r-i$ & $i-z$ & $r$ & $u-g$ & $g-r$ & $r-i$ & $i-z$ \\ \hline
    Scheme 1 & & &X &  &  & & &  &  &   \\ 
    Scheme 2 & X& & &  &  & X& &  &X  &   \\ 
    Scheme 3 & X& X& &  &  & & &  &  &    \\ 
    SDSS & X&X &  &  &  &X & X& X &  &    \\ 
    \end{tabular}
    \label{tab::variableSelecBeta}
\end{center}
\end{table}

\section{Conditional Density Estimation under Covariate Shift}
\label{sec-densityUnderCS}

\comment{In the statistical literature, conditional density estimation is considered to be difficult even for as few as 3 covariates \citep{Fan2}. This has motivated several works in settings without selection bias (e.g., \citealt{Hall2,Efromovich3,LeeIzbicki}).} 

Conditional density estimators are typically designed to minimize the loss 
\begin{align}
\label{eq::lossCDETrad}
\iint \left(\widehat{f}(z|\vec{x})-f(z|\vec{x})\right)^2 dP_L(\vec{x})dz 
\end{align}
under the implicit assumption that $\P_L=\P_U$. One can easily  estimate 
 this loss (up to a constant) using the labeled data:
\begin{align} \label{eq:empirical_loss_noshift}
 \frac{1}{\widetilde{n}_L}\sum_{k=1}^{\widetilde{n}_L}\int \widehat{f}^2\left(z|\widetilde{\vec{x}}_k^L\right)dz-2\frac{1}{\widetilde{n}_L}\sum_{k=1}^{\widetilde{n}_L}\widehat{f}\left(\widetilde{z}^L_k|\widetilde{\vec{x}}^L_k\right). 
\end{align}
However, our goal is       
 really to minimize
\begin{align}
\label{eq-lossCDE}
L(\widehat{f},f) :=\iint \left(\widehat{f}(z|\vec{x})-f(z|\vec{x})\right)^2dP_U(\vec{x})dz \, ,
\end{align}
where $\P_U$ is the distribution of the {\em unlabeled} target data. The two losses can be very different if $\P_L \neq \P_U$. Hence, a density estimator that performs well on the labeled data may not be a good estimator for the data of interest; Fig~\ref{fig::csExample} shows a similar problem in linear regression. 

  The challenge is to estimate Eq.~\ref{eq-lossCDE} without knowing $z$ at the unlabeled data points.
 Under the covariate shift assumption  $f_U(z|\vec{x})=f_L(z|\vec{x})$, one can use importance sampling. 
 Up to a constant, 
 Eq.~\ref{eq-lossCDE} becomes
\begin{align*}
L(\widehat{f},f) &=\iint \widehat{f}^2(z|\vec{x})dP_U(\vec{x})dz-2\iint \widehat{f}(z|\vec{x})f(z|\vec{x})dP_U(\vec{x})dz  \\
&= \iint \widehat{f}^2(z|\vec{x})dP_U(\vec{x})dz-2\iint \widehat{f}(z|\vec{x})\beta(\vec{x})dP_L(z,\vec{x}) \,,
\end{align*}
where we for the second equality use that $f_U(z|\vec{x})dP_U(\vec{x})dz = 
f_L(z|\vec{x}) \beta(x) dP_L(\vec{x})dz=\beta(x)dP_L(z,\vec{x})$. 
Hence, when  $\P_L \neq \P_U$, we replace the standard empirical loss in Eq.~\ref{eq:empirical_loss_noshift} by
\begin{align}
\label{eq-lossCDEEstimate}
\widehat{L}(\widehat{f},f) = \frac{1}{\widetilde{n}_U}\sum_{k=1}^{\widetilde{n}_U}\int \widehat{f}^2\left(z|\widetilde{\vec{x}}_k^U\right)dz-2\frac{1}{\widetilde{n}_L}\sum_{k=1}^{\widetilde{n}_L}\widehat{f}\left(\widetilde{z}^L_k|\widetilde{\vec{x}}^L_k\right)\widehat{\beta}\left(\widetilde{\vec{x}}^L_k\right).
\end{align}
We can compute this loss using (labeled and unlabeled) validation data and $\widehat{\beta}\left(\widetilde{\vec{x}}^L_k\right)$, which are estimates of the importance weights at the labeled data points.

 
 In what follows (Secs.~\ref{sec-NN}-\ref{sec-combine}), we present four conditional density estimators specifically designed for multivariate data and a covariate shift (CS) setting: \emph{NN}$_{CS}$, \emph{ker-NN}$_{CS}$, $Series_{CS}$, and Comb$_{CS}$.  The \emph{NN}$_{CS}$ nearest-neighbor histogram estimator first appeared in \citet{Cunha1}; the other estimators represent novel approaches.  
  For model selection and tuning of parameters, we minimize Eq.~\ref{eq-lossCDEEstimate} with importance weights $\beta$ estimated via Eqs.~\ref{eq-nnWeights} and \ref{eq-lossBetaEst}. The same loss is also used for model assessment of the SDSS data. 
For model assessment of the {\em simulated} data (where we know the 
true labels $\widetilde{z}^U$), we compute the more accurate error estimate 
\begin{align}
\label{eq-lossCDEEstimate2}
\widehat{L}(\widehat{f},f) = \frac{1}{\widetilde{n}_U}\sum_{k=1}^{\widetilde{n}_U}\int \widehat{f}^2\left(z|\widetilde{\vec{x}}_k^U\right)dz-2\frac{1}{\widetilde{n}_U}\sum_{k=1}^{\widetilde{n}_U}\widehat{f}\left(\widetilde{z}^U_k|\widetilde{\vec{x}}^U_k\right),
\end{align}
which does not involve importance weights $\beta$.


\begin{remark}[Choice of Loss Function]  
One can replace the averaged $L^2$-loss in Eq.~\ref{eq-lossCDE} with other measures of discrepancy, but many of the distance measures common in discrimination analysis (e.g. $F$-divergences and differences in log-densities) are profoundly sensitive to the tails of the distribution and not suitable for density estimation; see e.g. \citet{hall1987kullback} and \citet{wasserman}.
\end{remark}

\subsection{Nearest-Neighbor Histogram (NN$_{CS}$)}
\label{sec-NN}
In  \citet{Cunha1}, the authors use a weighted nearest-neighbor histogram to estimate the photo-$z$ distribution of a galaxy with photometric covariates $\x$.  Let $\mathcal{N}_N(\vec{x})$ denote the $N$ nearest neighbors of $\vec{x}$ among the labeled data. Divide $[0,1]$, the range of $z$, into $B$ equal-sized bins, 
and let $b(z)$ denote the bin that includes $z$ for $z \in [0,1]$. Then, the weighted histogram estimator is
\begin{align}
\label{eq-cunhann}
\widehat{f}(z|\vec{x})\propto \sum_{k\in \mathcal{N}_N(\vec{x})} \widehat{\beta}\left(\vec{x}^L_k\right) \I\left(z^L_k \in b(z)\right) \,,
\end{align}
where the importance weights $\widehat{\beta}(\vec{x}^L_k)$ reflect how representative each labeled galaxy $k\in \mathcal{N}_N(\vec{x})$ is of the 
target distribution. Cunha et al. choose the tuning parameters in their model by hand. Here we use Eq.~\ref{eq-lossCDEEstimate} to find the optimal values of $N$ and $B$ for the conditional density estimator in Eq.~\ref{eq-cunhann}, 
and we use Eq.~\ref{eq-lossBetaEst} to select the best value of $M$ in Eq.~\ref{eq-nnWeights} for computing the weights $\widehat{\beta}(\vec{x})$.

\subsection{Kernel Nearest-Neighbor Estimator (ker-NN$_{CS}$)}
\label{sec-ker-NN}

A simple way of improving upon $NN_{CS}$ is to replace the indicator function
in Eq.~\ref{eq-cunhann} with a kernel smoother:
\begin{align}
\label{eq::kernelNN}
\widehat{f}(z|\vec{x})\propto \sum_{k\in \mathcal{N}_N(\vec{x})}  \widehat{\beta}\left(\vec{x}^L_k\right) K_{\epsilon}\left(z-z^L_k\right) \,,
\end{align}
where, e.g., $K_{\epsilon}(z\!-\!z_k)\!=\!e^{-\left(z\!-\!z_k\right)^2\!/\!4\epsilon}$.\footnote{For a traditional kernel nearest neighbors 
estimator \emph{not} corrected for selection bias \citep{Lincheng}, let $\widehat{\beta}(\vec{x}) \equiv 1$ for all $\vec{x}$; 
we denote the uncorrected estimator by \emph{ker-NN}.} As before, we choose the tuning parameters   (here, $N$ and $\epsilon$ in Eq.~\ref{eq::kernelNN}, and $M$ in Eq.~\ref{eq-nnWeights}) that 
minimize the estimated losses in Eqs.~\ref{eq-lossCDEEstimate} and 
\ref{eq-lossBetaEst}. 
This estimator leads to much more accurate density estimates than its histogram equivalent when we have limited amounts of labeled data (and hence small values of $N$). 

\subsection{Spectral Series CDE under Covariate Shift (Series$_{CS}$)}
\label{sec-CDEunderCS}
Suppose that the covariates 
$\vec{x}$ lie in a lower-dimensional subspace $\mathcal{X}$ of
$\mathbb{R}^d$, where the number of covariates $d$ may be large. \citet{IzbickiLee_CDE} propose a spectral series estimator that expands the conditional density $f(z|\vec{x})$ in a basis $\Psi_{i,j}(z,\vec{x})=\phi_i(z)\psi_j(\vec{x})$ adapted to the {\em intrinsic} (lower-dimensional) geometry of a reference distribution $\P$ on $\mathcal{X}$. Here we generalize the series approach to a setting with covariate shift. We choose $\P_L$ as the reference distribution and tune the estimator so as to minimize the loss with respect to $\P_U$.

More specifically: We assume the functions $\phi_i$ to be standard (one-dimensional) Fourier 
basis functions, whereas the functions $\psi_j$ are the eigenfunctions of the 
operator 
$\textbf{K}: L^2(\mathcal{X},\P_L) \longrightarrow L^2(\mathcal{X},\P_L)$,
\begin{align}
\label{eq::kpcaOperator}
\textbf{K}(h)(\vec{x})=\int_\mathcal{X} K(\vec{x},\vec{y})h(\vec{y})dP_L(\vec{y}),
\end{align}
where $K(\vec{x},\vec{y})$ is a bounded, symmetric, and positive definite kernel. 
By construction \citep{Mink,IzbickiLee_CDE},
\begin{align}\label{eq:orthogonality}
\int_\mathcal{X} \psi_i(\vec{x})\psi_j(\vec{x})dP_L(\vec{x})  =\delta_{i,j}\overset{\mbox{\tiny{def}}}{=}\I(i=j).
\end{align} 
As a result, the coefficients in the series expansion are simply expectations over the eigenfunctions:
\begin{align}
\label{eq-coefficientCS}
\alpha_{i,j}&=\iint f(z|\vec{x}) \Psi_{i,j}(z,\vec{x})dP_L(\vec{x})dz=\E_{(\X,Z) \sim \P_L}[\Psi_{i,j}(Z,\vec{X})] \,.
\end{align}

In practice, we need to estimate both the functions  $\psi_j$  and the coefficients $\alpha_{i,j}$ from data:
Using the labeled training examples, we compute the first $J$ eigenvectors 
$\widetilde{\psi}_1,\ldots,\widetilde{\psi}_J$ of 
the Gram matrix
$$\left[ K\left(\vec{x}^L_i,\vec{x}^L_j\right) \right]_{i,j=1}^{n}, $$ 
where $K(\vec{x},\vec{y}) =
\exp{\left(-d^2(\vec{x},\vec{y})/4\epsilon\right)}$ is the Gaussian kernel. We then 
extend these vectors to out-of-sample points via the Nystr\"om extension
\begin{align}
 \label{eq::Nystrom}
 \widehat{\psi}_j(\vec{x})=\frac{\sqrt{n_L}}{\widehat{l}_j}\sum_{k=1}^{n_L} \widetilde{\psi}_j\left(\vec{x}^L_k\right) K\left(\vec{x},\vec{x}^L_k\right),
\end{align}
where $\widehat{l}_j$ is the eigenvalue associated to the eigenvector 
$\left(\widetilde{\psi}_j(\vec{x}^L_1),\ldots,\widetilde{\psi}_j(\vec{x}^L_{n_L})\right)$.   Next we estimate the expansion coefficients in Eq.~\ref{eq-coefficientCS} according to 
\begin{align*}
\widehat{\alpha}_{i,j} = \frac{1}{n_L}\sum_{k=1}^{n_L}\widehat{\Psi}_{i,j}(z_k^L,\vec{x}_k^L) \,,
\end{align*}
i.e., we average the empirical basis functions 
$\widehat{\Psi}_{i,j}(z_k,\vec{x}_k)=\phi_i(z_k)\widehat{\psi}_j(\vec{x}_k)$
over the labeled data. We define the new series estimator Series$_{CS}$ by
\begin{align}
\label{eq-finalEstimator}
\widehat{f}(z|\vec{x})=\sum_{i=1}^I\sum_{j=1}^J\widehat{\alpha}_{i,j}\widehat{\Psi}_{i,j}(z,\vec{x}) \, ,
\end{align}
where the parameters $I$, $J$, and $\epsilon$ are chosen so as to minimize the 
loss in Eq.~(\ref{eq-lossCDEEstimate}) relative to the {\em unlabeled} data $P_U$.

\subsection{Combined Estimator (Comb$_{CS}$)}
\label{sec-combine}
Finally, we present a procedure for combining, or {\em stacking}, multiple estimators in a principled way. Suppose that $\widehat{f}_1(z|\vec{x}),\ldots, \widehat{f}_p(z|\vec{x})$ are different  estimators of $f(z|\vec{x})$; these estimators could, for example, be any of the cross-validated estimators described in Secs.~\ref{sec-NN}-\ref{sec-CDEunderCS}. Now ask the question: Can we average these models so as to reduce the prediction performance of individual estimators? If we restrict ourselves to weighted averages, then the answer is to compute
\begin{align*}
 \widehat{f}^{{\boldsymbol \alpha}}(z|\vec{x})=\sum_{k=1}^p \alpha_k \widehat{f}_k(z|\vec{x}),
\end{align*}
where the weights minimize the empirical loss
$\widehat{L}(\widehat{f}^{{\boldsymbol \alpha}},f)$ in Eq.~\ref{eq-lossCDEEstimate} under the constraints $\alpha_i \geq 0$ and $\sum_{i=1}^p \alpha_i=1$.  The weights ${\boldsymbol \alpha}=\left[\alpha_i \right]_{i=1}^p$ can then be found by solving a standard quadratic programming problem: 
\begin{align}
\label{eq-quadratic}
 \underset{{\boldsymbol \alpha}:\alpha_i\geq 0,\sum_{i=1}^p \alpha_i=1}{\arg\min} {\boldsymbol \alpha}' \mathbb{B} {\boldsymbol \alpha} -2{\boldsymbol \alpha}'b ,
\end{align}
where $\mathbb{B}$ is the $p \times p$ matrix $\left[\frac{1}{\widetilde{n}_U} \sum_{k=1}^{\widetilde{n}_U} \int \widehat{f}_i(z|\widetilde{\vec{x}}_k^U)\widehat{f}_j(z|\widetilde{\vec{x}}_k^U)dz\right]_{i,j=1}^p$
and $b$ is the vector
$\left[ \frac{1}{\widetilde{n}_L}\sum_{k=1}^{\widetilde{n}_L}\widehat{f}_i(\widetilde{z}^L_k|\widetilde{\vec{x}}^L_k)\widehat{\beta}(\widetilde{\vec{x}}^L_k)\right]_{i=1}^p.$ 

 In this work, \emph{Comb$_{CS}$} denotes the estimator that combines the two estimators  \emph{ker-NN$_{CS}$} and Series$_{CS}$, although this procedure of combining models applies more generally to other conditional density estimators.

\subsection{Experiments}\label{sec:CDE_experiments}
\label{sec-analyses}
Using the simulated and observed
data (Sec.~\ref{sec-data}), we will now compare the performance of seven different estimators of $f(z|\x)$.  (As before, our covariates 
are the four colors and the $r$-band magnitude in the {\tt model} 
magnitude system. We split our data into training, validation, and test sets as in Sec.~\ref{sec::appliWeights}.) The first three estimators {\em do not account for selection bias}. They are
\begin{itemize}
\item \emph{NN}: the nearest neighbor estimator from Sec.~\ref{sec-NN} with $\widehat{\beta}(\vec{x}) \equiv 1$ for all $\vec{x}$;
\item \emph{ker-NN}: the kernel nearest neighbor estimator from Sec.~\ref{sec-ker-NN} with  $\widehat{\beta}(\vec{x}) \equiv 1$ for all $\vec{x}$; and
\item \emph{Series}: the spectral series estimator from  \cite{IzbickiLee_CDE}.
\end{itemize}
The tuning parameters of these three estimators minimize the 
empirical loss in Eq.~\ref{eq:empirical_loss_noshift} on the (labeled) validation data. The
second three estimators {\em correct for covariate shift by importance weights}. They are
\begin{itemize}
\item \emph{NN$_{CS}$}: the nearest neighbor estimator from Sec.~\ref{sec-NN};
\item \emph{ker-NN$_{CS}$}: the kernel nearest neighbor estimator from 
Sec.~\ref{sec-ker-NN}; and 
\item \emph{Series$_{CS}$}: the spectral series estimator from
 Sec.~\ref{sec-CDEunderCS}. 
\end{itemize}
Finally, we have
\begin{itemize}
\item \emph{Comb$_{CS}$}: an estimator that 
combines \emph{ker-NN$_{CS}$} and \emph{Series$_{CS}$} according to Sec.~\ref{sec-combine}. 
\end{itemize}
We choose the tuning parameters of these last four estimators so as to minimize the reweighted empirical loss $\widehat{L}(\widehat{f},f)$
 in Eq.~\ref{eq-lossCDEEstimate} on (labeled and unlabeled) validation data. By bootstrap, we estimate the standard error of  $\widehat{L}(\widehat{f},f)$ according to
$\sqrt{\V\left[\widehat{L}(\widehat{f},f)\right]} \approx \sqrt{B^{-1}\sum_{b=1}^B \left(\widehat{L}_b(\widehat{f},f)-\overline{\widehat{L}(\widehat{f},f)} \right)^2},$
where $B=500$ is the number of bootstrap samples of the test set, $\widehat{L}_b(\widehat{f},f)$ is the estimated loss for the $b$th bootstrap sample, and
$\overline{\widehat{L}(\widehat{f},f)}$ is the mean of 
$\{\widehat{L}_b(\widehat{f},f)\}_{b=1}^B$.
Note that each bootstrap sample consists of a sample with replacement from the labeled set and a  sample with replacement from the unlabeled set.

\begin{figure}[H]
  \centering
\subfloat[Scheme 1]{\includegraphics[trim={0 1.7cm 0 0},clip,page=1,scale=0.15]{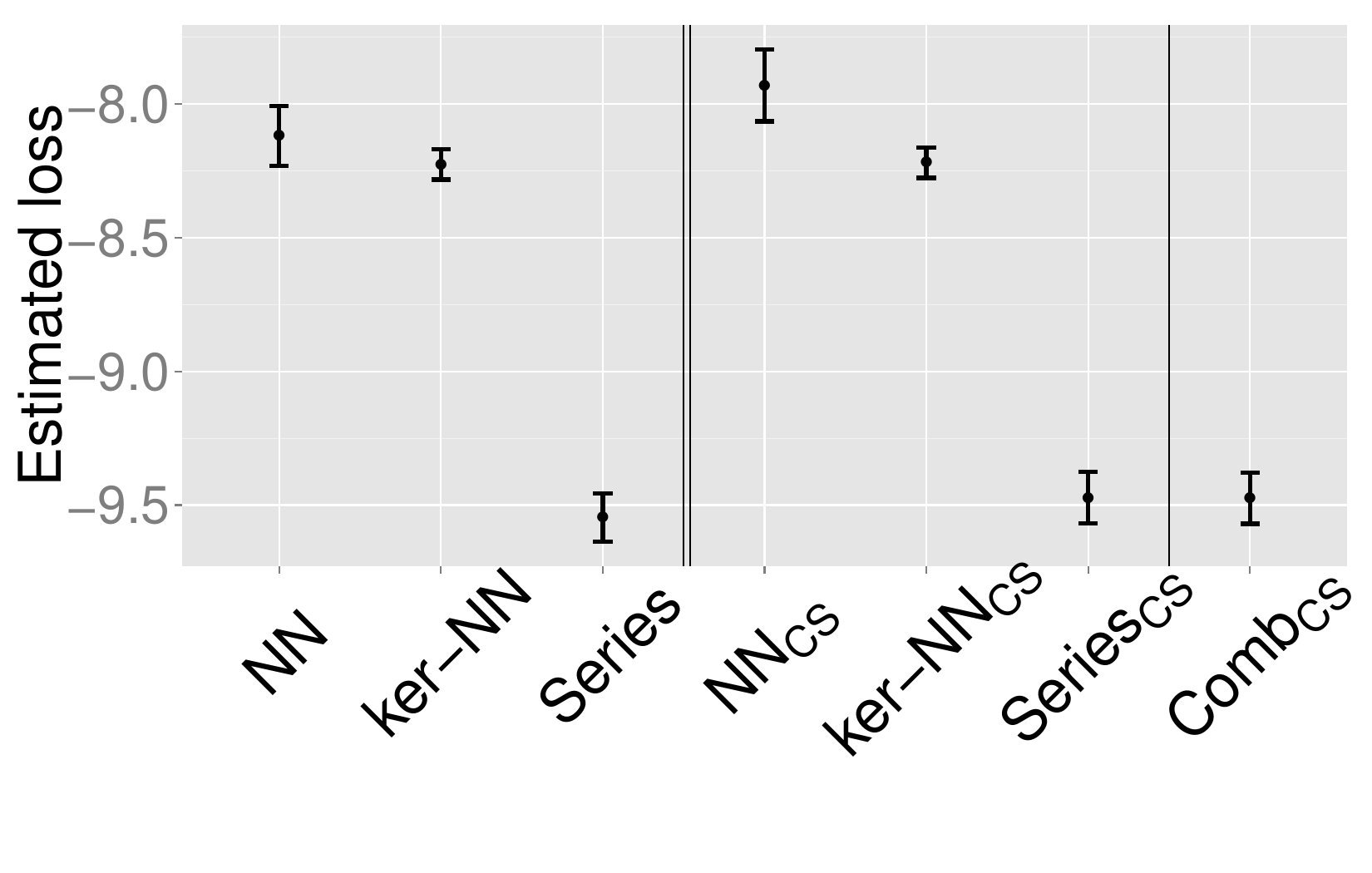}}
\subfloat[Scheme 2]{\includegraphics[trim={0 1.7cm 0 0},clip,page=2,scale=0.15]{GGPLOTFinalEstimatorsSingleSplit.pdf}}\\[-0.6mm]
\subfloat[Scheme 3]{\includegraphics[trim={0 1.7cm 0 0},clip,page=3 ,scale=0.15]{GGPLOTFinalEstimatorsSingleSplit.pdf}}
\vspace{-1mm}
\caption{\small Estimated loss of different density estimators for Schemes 1-3 with, respectively, (a) no, (b) moderate, and (c) large covariate shift. Bars correspond to mean plus and minus standard error.  \emph{ker-NN} is robust to selection bias and has a much smaller loss than previously proposed NN estimators (\emph{NN} and \emph{NN$_{CS}$}) even without importance weights; see text for discussion.
\label{fig::riskSDSSLabeled5}}
\end{figure}

\begin{figure}[H]
  \centering
\subfloat[{\footnotesize Original Sample}]{\label{fig::finalestimatesRealB}\includegraphics[trim={0 1.7cm -2cm 0},clip,page=1,scale=0.195]{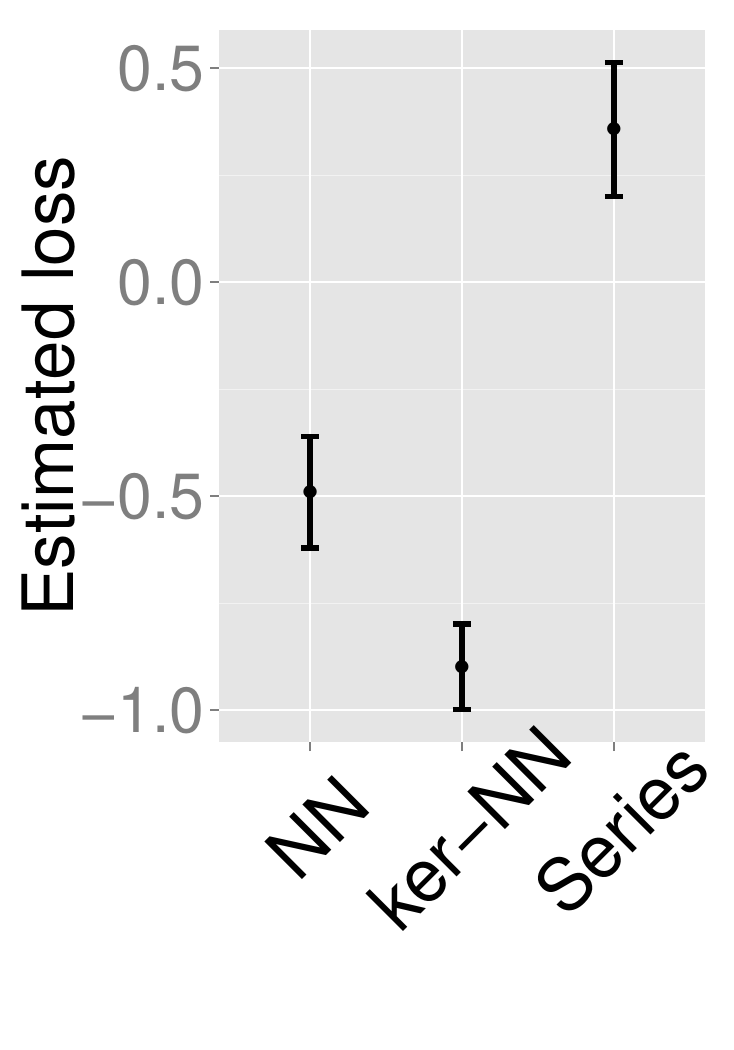}}\hspace{4mm}
\subfloat[{\footnotesize After removing $\widehat{\beta}(\vec{x})=0$}]{\label{fig::finalestimatesRealA}\includegraphics[trim={0 1.7cm 0 0},clip,page=1,scale=0.195]{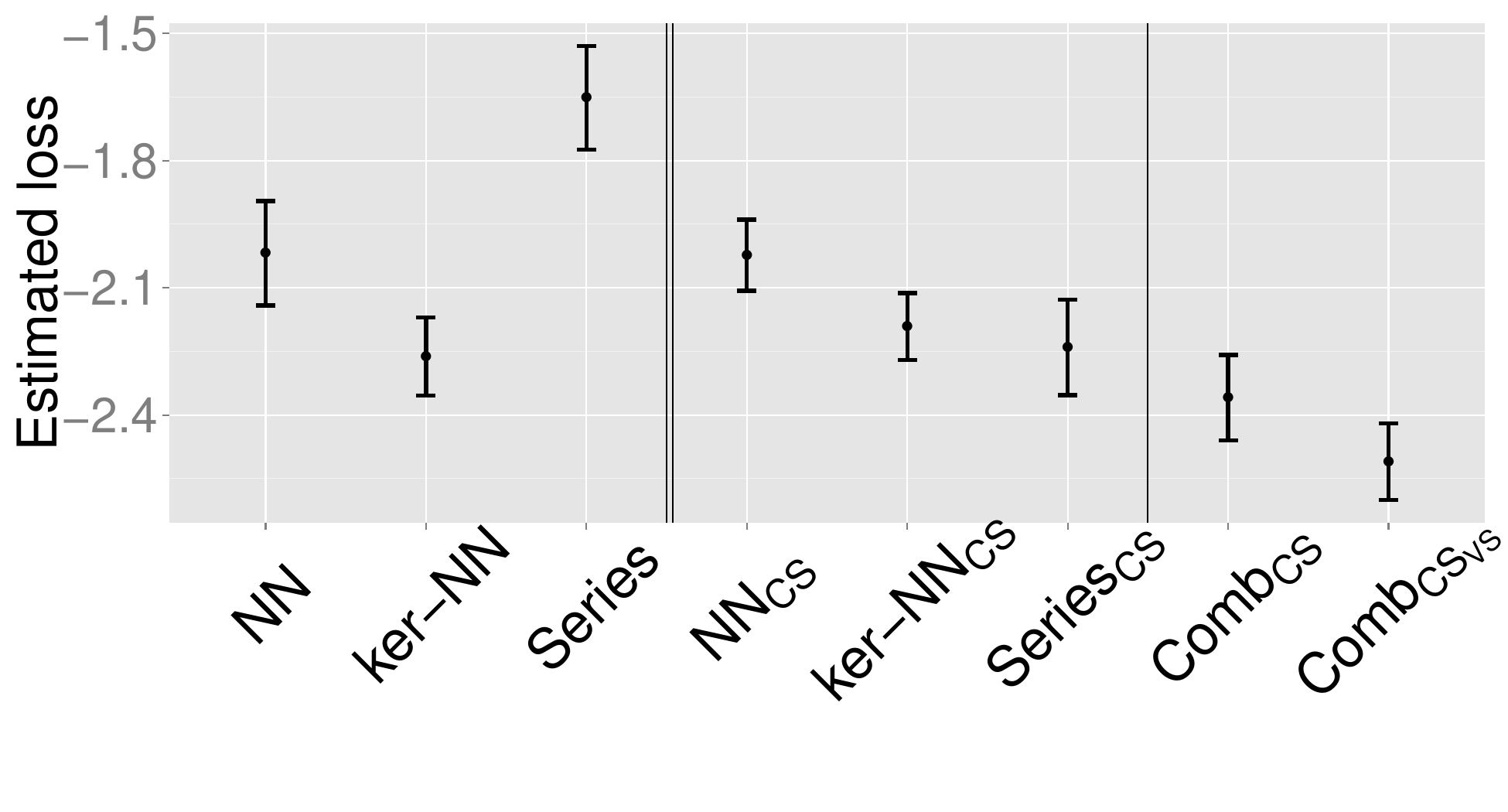}}
  \caption{\small Estimated loss of conditional density estimators for SDSS data 
      (a) in  the original sample with no preprocessing, and (b) after removing and replacing labeled examples for which the initial weight estimates $\widehat{\beta}(\x)=0$.
  Note that the scales in the two plots differ.}
  \label{fig::finalestimatesReal}
\end{figure}

Fig.~\ref{fig::riskSDSSLabeled5} shows the empirical loss 
 (Eq.~\ref{eq-lossCDEEstimate2}) on the test set 
for the simulated samples with known covariate shifts. Fig.~\ref{fig::finalestimatesRealB} shows the 
estimated loss (Eq.~\ref{eq-lossCDEEstimate}) for 15,000 SDSS samples with no preprocessing, 
and Fig.~\ref{fig::finalestimatesRealA} shows the loss after 
removing and replacing training examples with estimated weights $\widehat{\beta}(\x)\!=\!0$ so as to increase the effective sample size; note that the scales in these two plots differ.
  \vspace{-7mm}

{\em Variable Selection.}  As in Sec.~\ref{sec::appliWeights}, one can further improve these results 
 by choosing a subset of the ten covariates from the {\tt model} 
and {\tt cmodel} magnitude systems. Table \ref{tab::variableSelecCDE} 
lists the covariates from a 
forward stepwise model search with the combined estimator, initialized by an estimate of the 
marginal distribution $f(z)$. For the SDSS data, the loss 
of the combined estimator with variable selection (\emph{Comb$_{{CS}_{VS}}$} 
in Fig.~\ref{fig::finalestimatesRealA}) is $-2.51 \pm 0.09$, which is 
smaller than $-2.36 \pm 0.10$, the loss achieved by the combined estimator 
(\emph{Comb$_{CS}$}) based on a fixed set of five {\tt model} covariates.

\comment{Finally, Fig.~\ref{fig::estimates} shows examples 
of conditional density estimates 
for the SDSS data using \emph{Series$_{CS}$}, \emph{ker-NN$_{CS}$} and 
\emph{Comb$_{CS}$}. Most estimates are multimodal and asymmetric, 
indicating that the regression $\E[Z|\x]$ may not adequately describe the 
association between $\x$ and $Z$.}
  \vspace{-3mm}

\comment{
\begin{figure}[H]
  \centering
\subfloat{\includegraphics[page=1,scale=0.34]{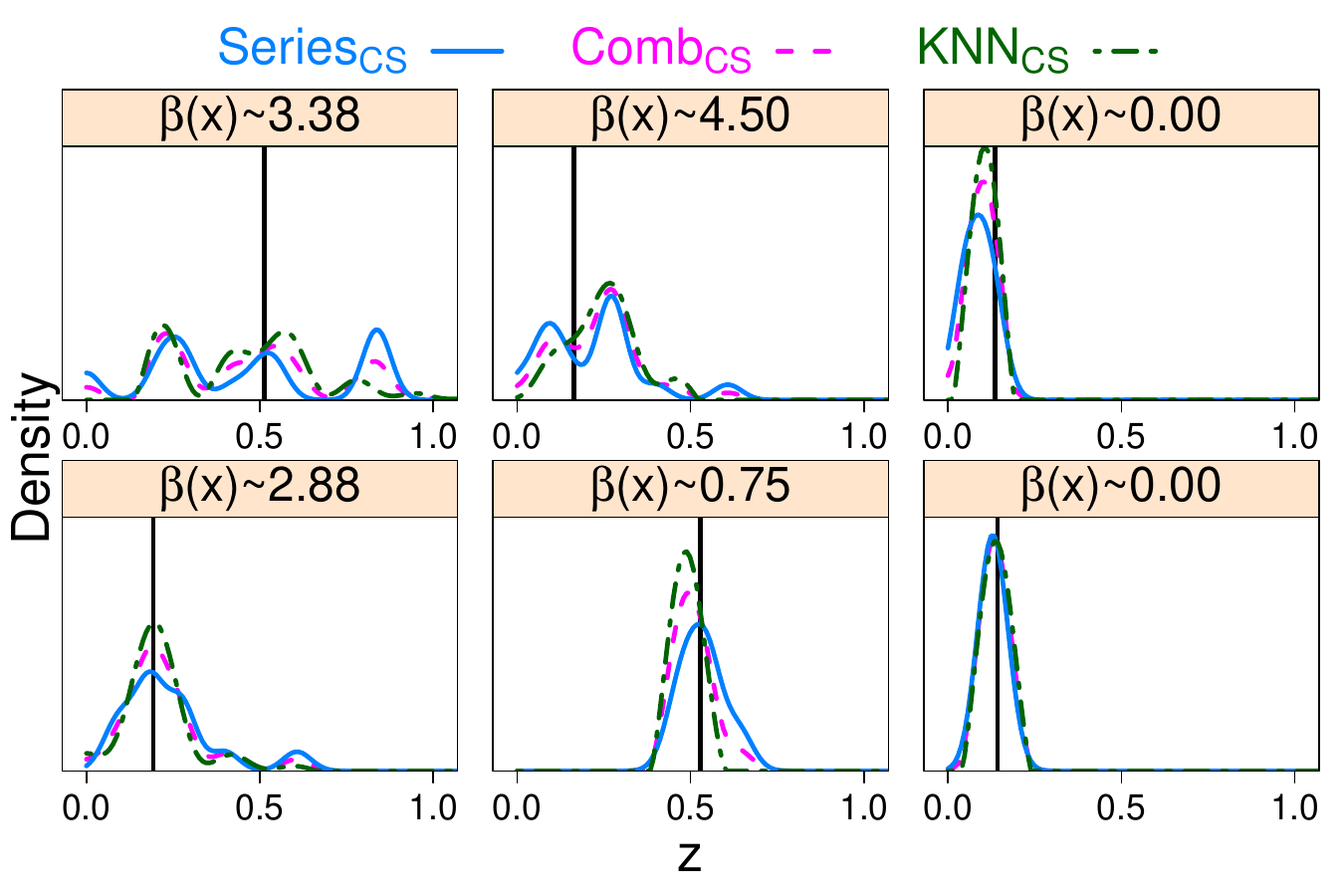}}\\[-3mm]
\subfloat{\includegraphics[page=1,scale=0.34]{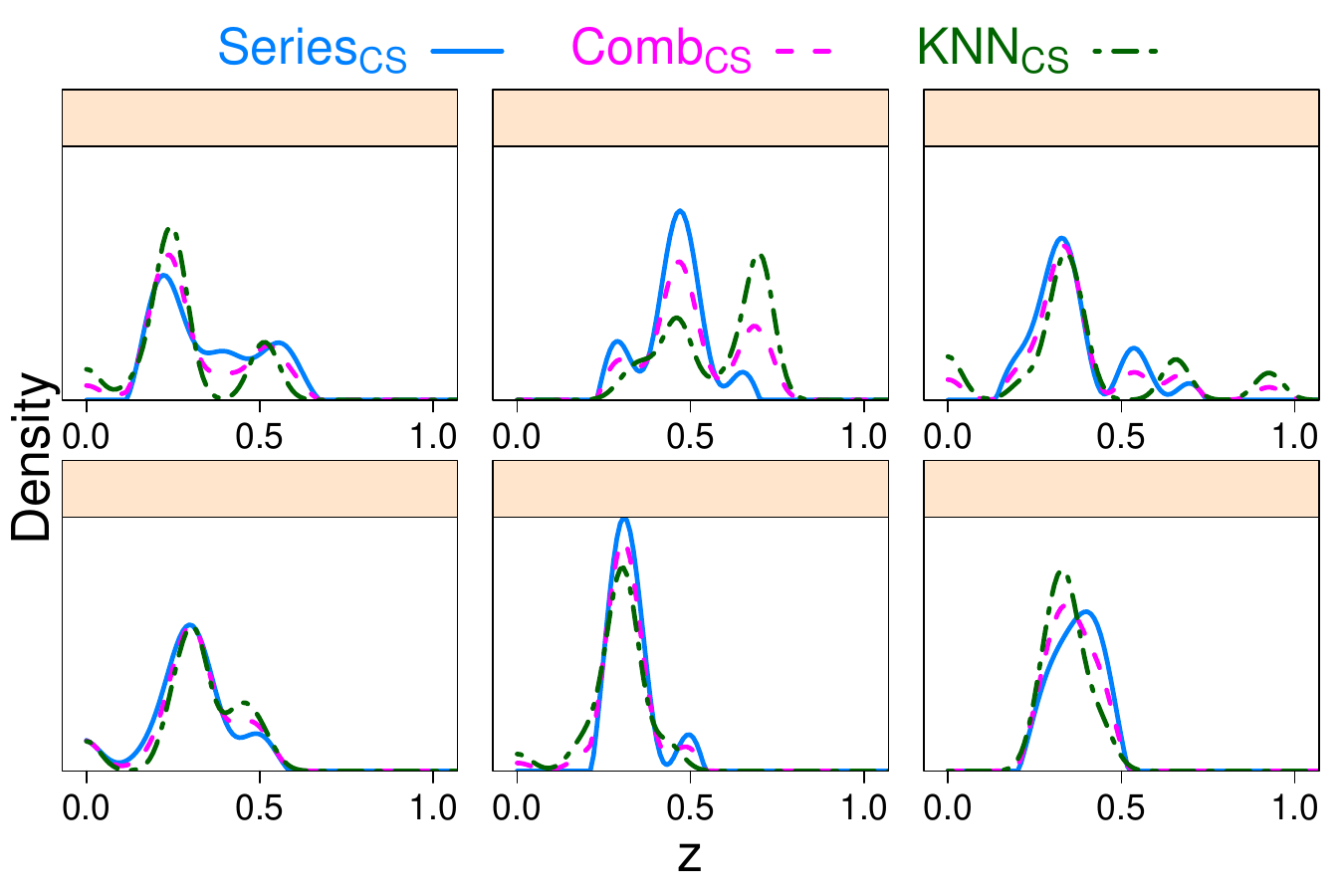}}
  \caption{\small {\em Top}: Estimated densities and estimated importance weights for six galaxies in the spectroscopic sample. Vertical lines indicate the true (spectroscopically confirmed) redshift.
  {\em Bottom}: Examples of estimated densities for six galaxies in the photometric sample.}
    \label{fig::estimates}
\end{figure}
}
\vspace{-6mm}

\begin{table}[H]
\begin{center}
\caption{\small Selected covariates for conditional density estimation with \emph{Comb$_{{CS}_{VS}}$}}
    \begin{tabular}{  l | c |c |c |c |c |c |c |c |c |c }
    \hline
    Data set &  \multicolumn{5}{c|}{{\tt model}}  &  \multicolumn{5}{c}{{\tt cmodel}}  \\ \hline
    & $r$ & $u-g$ & $g-r$ & $r-i$ & $i-z$ & $r$ & $u-g$ & $g-r$ & $r-i$ & $i-z$ \\ \hline
    Scheme 1 & X&X & X& X& X& && X& X& X \\ 
    Scheme 2 & X& &X &X & & X& & & &  \\ 
    Scheme 3 & X& &X &X & &X & & & &  \\ 
    SDSS &X & &X &X & &X & & & &  \\ 
    \end{tabular}
    \label{tab::variableSelecCDE}
\end{center}
\end{table}

{\noindent \em Summary.} Our main conclusions are as follows.
\begin{enumerate} 
\item A necessary condition for importance weighting is that $\mathbb{P}_L$ dominates $\mathbb{P}_U$. Our results (Fig.~\ref{fig::finalestimatesReal}) show that one should {\em also} restrict the labeled data to regions where there is unlabeled data; a simple procedure is to search for labeled examples with $\widehat \beta(\x)=0$ and replace these data with {\em new} labeled examples with $\widehat \beta(\x)\neq 0$. 

\item Our kernel-based estimators \emph{ker-NN} and \emph{ker-NN$_{CS}$} consistently perform better 
than their histogram counterparts \emph{NN} and \emph{NN$_{CS}$} introduced by \citet{Cunha1}.

\item Interestingly, \emph{ker-NN} is robust to selection bias even without explicitly incorporating importance weights into the estimator. 
Here is an intuitive explanation: If the neighborhood $N(\vec{x}_0)$ around $\vec{x}_0$ is sufficiently 
small, then the covariate shift 
assumption (Eq.~\ref{eq:covariate_shift}) implies that 
$f_L[z|\vec{x}\in N(\vec{x}_0)] \approx f_U[z|\vec{x}\in N(\vec{x}_0)]$.
As a result, \emph{ker-NN} returns good density estimates 
even without correcting for selection bias.  \emph{NN} is less robust than \emph{ker-NN} because smoothing via
binning requires larger neighborhoods (over which the above approximation may not hold) than
smoothing via kernels. For example, in Scheme 3, $N$ = 35 for \emph{NN},
versus 8 for \emph{ker-NN}.

\item \emph{Series} is sensitive to selection bias, but its covariate 
shift-corrected analogue \emph{Series$_{CS}$}, which we introduce in this work, is one of the best estimators for conditional density estimation.
When there is {\em no} selection bias (Fig.~\ref{fig::riskSDSSLabeled5}a), spectral series (\emph{Series}  and \emph{Series$_{CS}$}) perform significantly better than nearest neighbors methods (\emph{ker-NN} and \emph{ker-NN$_{CS}$}); this result is consistent with earlier work by \cite{IzbickiLee_CDE} on \emph{Series}. In settings {\em with} covariate shift (Fig.~\ref{fig::riskSDSSLabeled5}b-c and Fig.~\ref{fig::finalestimatesReal}b),  \emph{Series$_{CS}$} and \emph{ker-NN$_{CS}$} are comparable. We conjecture that \emph{Series$_{CS}$} may lose some of its competitive edge when $\P_L \neq \P_U$ because we compute the eigenvectors in the series using only labeled data, and then extrapolate to regions of the unlabeled data via the Nystr\"om extension (Eq.~\ref{eq::Nystrom}).

\item However, by {\bf combining} \emph{ker-NN$_{CS}$} and \emph{Series$_{CS}$} as in Sec.~\ref{sec-combine}
 we automatically get ``the best of both worlds'' under a variety of different settings. (The combined estimator \emph{Comb$_{CS}$} assigns weights $\alpha \!=\! 0.96$, 0.50, and 0.43 to \emph{Series$_{CS}$} in Schemes 1-3, and  $\alpha\!=\!0.53$ for the SDSS data.)  

\item Finally, we can further improve our predictions by {\bf variable selection} according to Secs. \ref{sec::appliWeights} and \ref{sec:CDE_experiments}
 when estimating $\beta(\x)$ and $f(z|\x)$, respectively. Selection bias leads to smaller models with fewer variables (see Table~\ref{tab::variableSelecCDE}), which is consistent with covariate shift decreasing the effective sample size
\citep{Shimodaira2000, Gretton},
 and thereby, increasing the variance of the estimators. 
\end{enumerate}

Algorithm~1 summarizes the combined model with variable selection (\emph{Comb$_{{CS}_{VS}}$}). This model includes two main steps, where we (i) first estimate the importance weights via $\beta$-NN with variable selection (lines 1-9), and (ii) then estimate the conditional density with the combined estimator with variable selection (lines 10-23) that combines Series$_{CS}$ (lines 11-14) and ker-NN$_{CS}$ (lines 16-20).

\begin{algorithm}[H]
	\caption{ \small Redshift estimation under covariate shift. The combined model with variable selection  (\emph{Comb$_{{CS}_{VS}}$}).}\label{algorithmCondReg}
	\begin{algorithmic}[1]
	\Require {\small  Labeled training data
			$(\x_1^L,z_1^L),\ldots,(\x_{n_L}^L,z_{n_L}^L)$; 
			labeled validation data
			$(\tilde{\x}_1^L,\tilde{z}_1^L),\ldots,
			(\tilde{\x}_{\tilde{n_L}}^L,\tilde{z}_{\tilde{n_L}}^L)$;  unlabeled training data
			$\x_1^U,\ldots,\x_{n_U}^U$; 
			unlabeled validation data
			$\tilde{\x}_1^U,\ldots,
			\tilde{\x}_{\tilde{n_U}}^U$; grid over tuning parameters $\epsilon_1$, $\epsilon_2$, $I$, $J$, $M$, $N$. }
	\Ensure {\small Estimator $\widehat{f}(z|\vec{x})$} 
	\item[]
  		\ForAll{$S \subset \{x_1,\ldots,x_d\}$}  \Comment{Alternatively, one may use a stepwise approach} 
	  		\ForAll{$M$}   \Comment{Tune NN Importance Weights Estimator using covariates $S$}
    		    \State Fit $\widehat{\beta}_{M,S}$  \Comment{Eq.~\ref{eq-nnWeights}}
		        \State Estimate $L(\widehat{\beta}_{M,S},\beta)$ \Comment{Eq.~\ref{eq-lossBetaEst}} 
        	\EndFor
        	\State Define $M^*=\arg \min_{M} \widehat{L}(\widehat{\beta}_{M,S},\beta)$.
	\State Let $\widehat{\beta}_{S}:=\widehat{\beta}_{M^*,S}$.
        \EndFor
  		\State Let $\widehat{\beta}:=\widehat{\beta}_{S^*}$, where $S^*=\arg \min_{S} L(\widehat{\beta}_{S},\beta)$
  		\item[] 
  		\ForAll{$S \subset \{x_1,\ldots,x_d\}$}  \Comment{Alternatively, one may use a stepwise approach} 
  			\ForAll{$I,J,\epsilon_1$}  \Comment{Tune Spectral Series Estimator using covariates $S$}
		  		\State Fit $\widehat{f}_{I,J,\epsilon_1,S}^{\mbox{series}}$ \Comment{Eq.~\ref{eq-finalEstimator}}
		  		\State Estimate $L(\widehat{f}_{I,J,\epsilon_1,S}^{\mbox{series}},f)$ \Comment{Eq.~\ref{eq-lossCDEEstimate}}
        	\EndFor  		 
	        \State Let $(I^*,J^*,\epsilon_1^*)=\arg \min_{I,J,\epsilon_1} L(\widehat{f}_{I,J,\epsilon_1,S}^{\mbox{series}},f)$ 
	          		\item[] 
    	    \ForAll{$N,\epsilon_2$}  \Comment{Tune Kernel Nearest-Neighbor Estimator using covariates $S$}
        		\State Fit $\widehat{f}_{N,\epsilon_2,S}^{\mbox{ker-NN}}$ \Comment{Eq.~\ref{eq::kernelNN}}
	        	\State Estimate $L(\widehat{f}_{N,\epsilon_2,S}^{\mbox{ker-NN}},f)$ \Comment{Eq.~\ref{eq-lossCDEEstimate}}
	        \EndFor  		
    	    \State Let $(N^*,\epsilon_2^*)=\arg \min_{N,\epsilon_2} L(\widehat{f}_{N,\epsilon_2,S}^{\mbox{ker-NN}},f)$ 
    	      		\item[] 
    	    \State Find $\widehat{f}_S$ by combining $\widehat{f}_{I^*,J^*,\epsilon_1^*,S}^{\mbox{series}}$ and $\widehat{f}_{N^*,\epsilon_2^*,S}$  \Comment{Sec.~\ref{sec-combine}}
    	    \State Estimate $L(\widehat{f}_S,f)$ \Comment{Eq.~\ref{eq-lossCDEEstimate}}
        \EndFor  		
          		\item[] 
        \State Output $\widehat{f}_{S^*}$, where $S^*=\arg \min_{S} L(\widehat{f}_{S},f)$
%
	\end{algorithmic}
\end{algorithm}

\begin{figure}[H]
  \centering
  \subfloat[\emph{NN$_{CS}$}]{\includegraphics[page=1,scale=0.21]{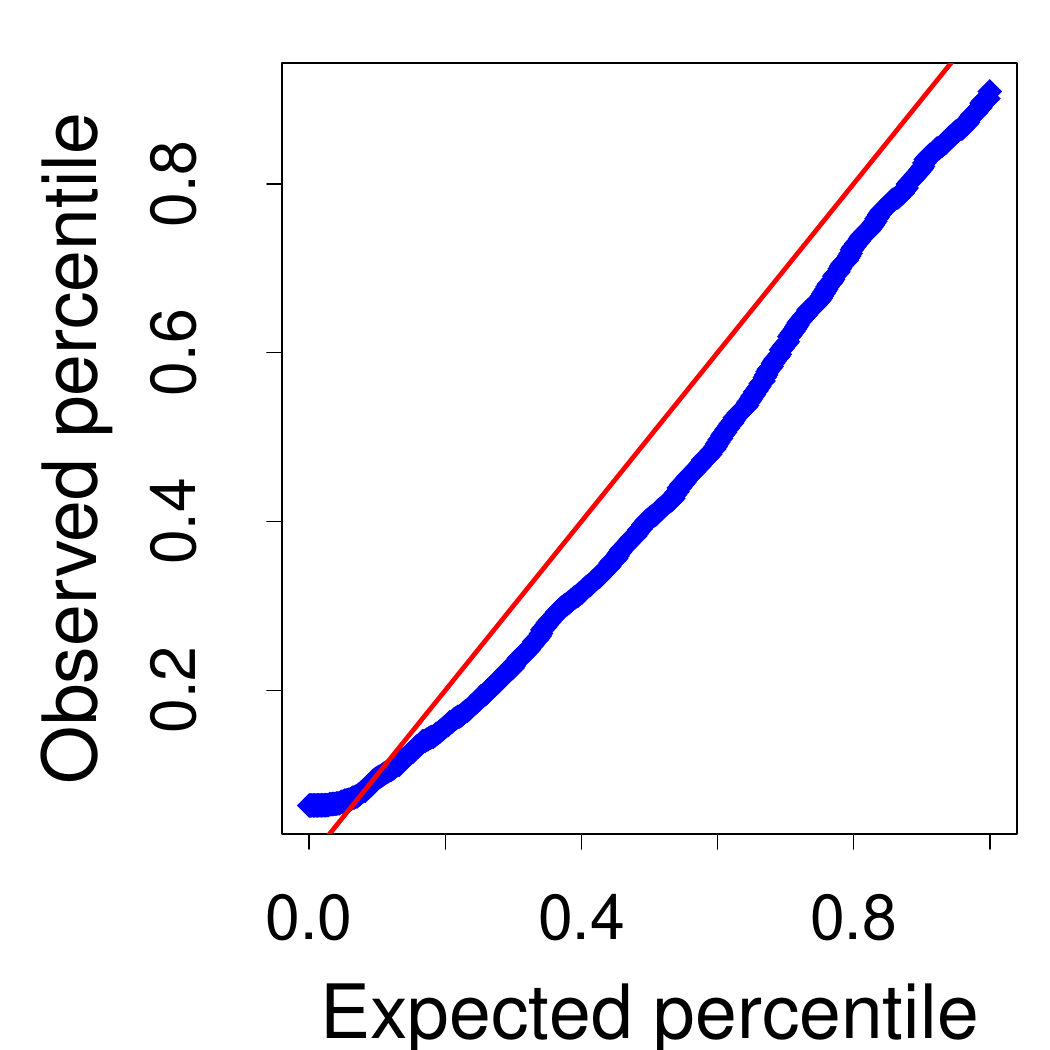}}\hspace{1mm}
\subfloat[\emph{Comb$_{{CS}_{VS}}$}]{\includegraphics[page=1,scale=0.21]{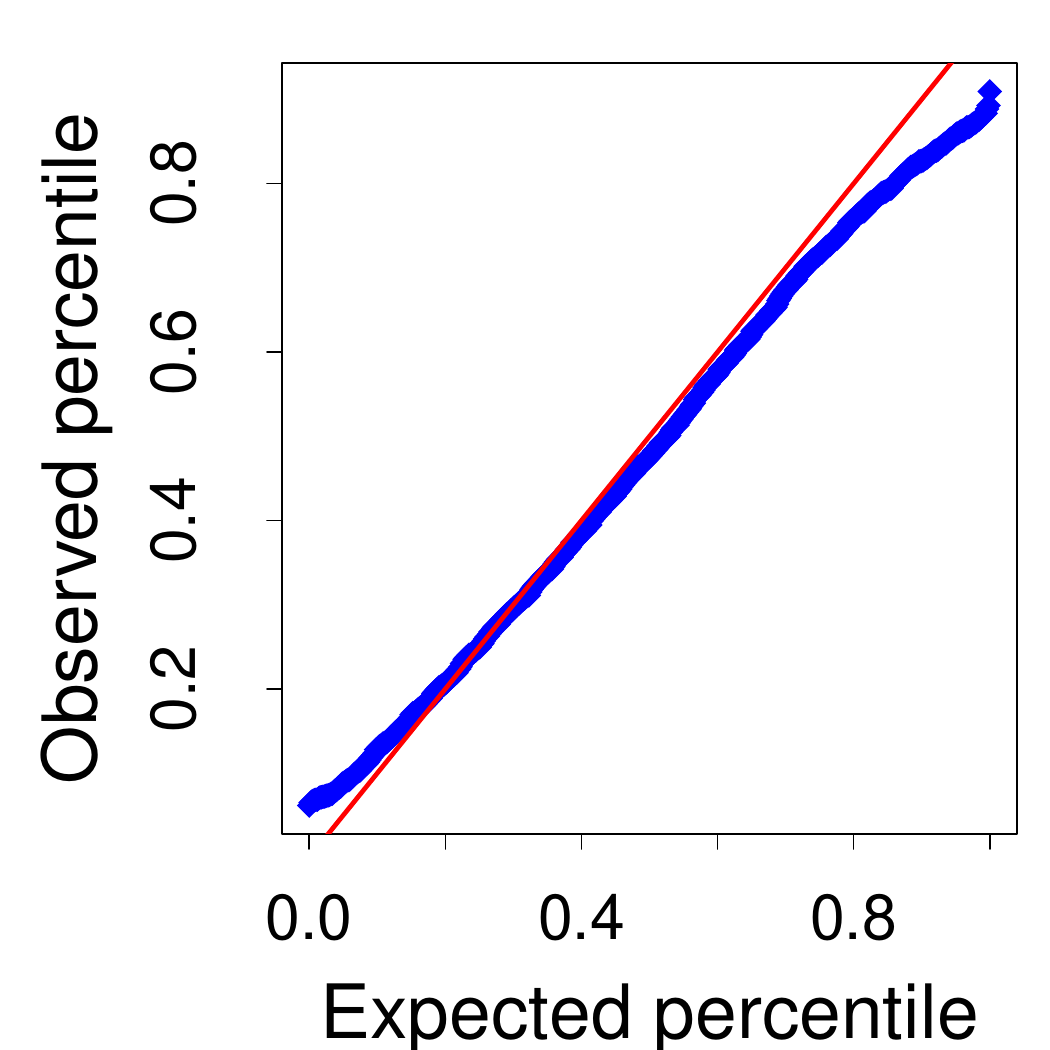}}\hspace{1mm}
  \caption{  Quantile-quantile plots for (a) \emph{NN$_{CS}$}  and (b)   \emph{Comb$_{{CS}_{VS}}$}, the combined model with selected covariates.}
  \label{fig::diag}
\end{figure}

In Appendix~\ref{app:diag}, we describe different diagnostic tests that can be used to more closely assess the quality of different models. Fig.~\ref{fig::diag}, for example, shows quantile-quantile (Q-Q) plots of the SDSS data for \emph{NN$_{CS}$} and  \emph{Comb$_{{CS}_{VS}}$}. These plots tell us how well these density estimates actually fit the observed data.
\vspace{-5.9mm}

\section{Application to Galaxy-Galaxy Lensing}
\label{sec-lensing}

By working with a probability distribution of the photometric redshift instead of a single best estimate, one can  reduce systematic biases in cosmological analyses (\citealt{mandelbaum,Wittman,Sheldon}). In this section, we study the galaxy-galaxy weak-lensing application from \cite{Sheldon} and
 offer new insights on the proper use of redshift distributions in downstream analysis; i.e., when estimating {\em functions} $g(z)$ of an unknown redshift $z$. \comment{Even though we ultimately are interested in $g(z)$, there are clear advantages in estimating the photo-$z$ density $f(z|\vec{x})$ well --- rather than just aiming for the best regression of $g(Z)$ on the photometric covariates $\x$. For example, the Sheldon et al data set includes $\approx 500{,}000$ lenses that each has a different function $g$, and future data sets will only increase in size. With a good density estimator $\widehat{f}(z|\vec{x})$ , one can address several different inference problems with different $g$'s simultaneously as well as construct reliable predictive intervals for each function $g(Z)$ for new observations of $Z$. }

Weak gravitational lensing is the slight deflection of photons from distant
astronomical sources that occurs when they pass near massive 
``lenses" (e.g., galaxies or galaxy clusters) lying closer to us.
Lensing acts to magnify the sizes of the distant sources as well as
to distort their shapes. 
Cosmologists can use it to directly probe the distribution of dark matter,
a form of matter that does not interact with light (hence the moniker
``dark") and which comprises $\approx$27\% of the mass-energy
of the Universe.
The \textit{critical surface density} $\Sigma(z_l,z_s)$ determines the lensing
strength of a given lens-source pair \citep{mandelbaum}.  The goal in this example is to estimate $g_l(z_s) = \Sigma^{-1}(z_l,z_s)$ for a 
  source galaxy with {\em unknown} redshift $z_s$, 
 assuming that the redshift $z_l$ of lens $l$ is known. 
A naive approach is to evaluate the function $g_l(\cdot)$ at a point estimate of the source galaxy redshift $z_s$; i.e., to compute $g_l(\widehat{z}_s) = \Sigma^{-1}(z_l,\widehat{z}_s)$ where $\widehat{z}_s$ typically is an estimate of the regression ${\E}[Z|\x]$.
 However, because
  $\E[g_l(Z)|\x] \neq g_l(\E[Z|\x])$, 
  the estimator
\begin{equation}\label{eq:lensing_calibration}
 \widehat{\Sigma^{-1}}(z_l,z_s) := \int \Sigma^{-1}(z_l,z) \widehat{f}(z|\vec{x})dz
\end{equation} 
usually  yields better results if $\widehat{f}(z|\vec{x})$ is a good estimate of $f(z|\x)$ \citep{Sheldon}.  Note that even though we ultimately are interested in $g_l(z)$, there are clear advantages in estimating the photo-$z$ density $f(z|\vec{x})$ well --- rather than just aiming for the best regression of $g_l(Z)$ on the photometric covariates $\x$. For example, the data set in this example includes $\approx 500{,}000$ lenses that each has a different function $g_l$, and future data sets will only increase in size. With a good density estimator $\widehat{f}(z|\vec{x})$, one can address several different inference problems with different $g_l$'s simultaneously as well as construct reliable predictive intervals for each function $g_l(Z)$ for new observations of $Z$.



Following \citet{Sheldon}, we 
  use data 
  from the DEEP2 EGS Region \citep{weiner2005deep}. In addition to the conditional density estimators in Sec.~\ref{sec-densityUnderCS}, we implement
 a nearest-neighbors density estimator with 7 neighbors (the value hand-picked by  \citealt{Sheldon} for this particular application), and \emph{photoz},  which computes $\Sigma^{-1}(z_l, \widehat{z}_s)$
 where $\widehat{z}_s$ is the nearest neighbor {\em regression} estimate of $z_s$. 
 To assess the performance of these methods in galaxy-galaxy lensing, we use the two measures in \citet{mandelbaum} called the lensing calibration bias and the variance ratio. 
Small values of bias and large values of variance ratio indicate good performance. We use 500 samples for training, 500 for validation, and 
382 for testing.

\begin{figure}[H]
  \centering
\subfloat[Bias]{\includegraphics[page=1,scale=0.21]{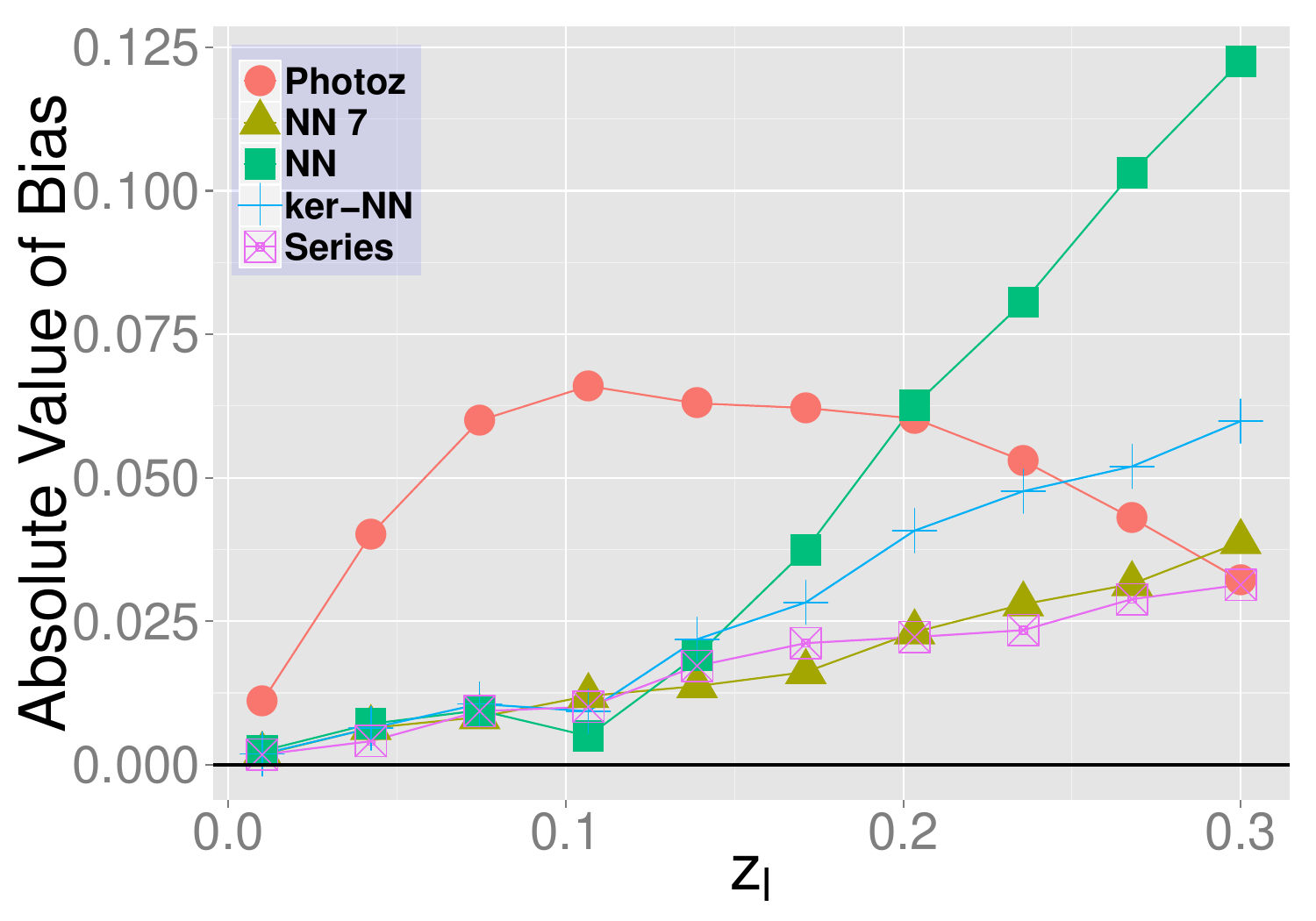}}
\subfloat[Variance ratio]{\includegraphics[page=2,scale=0.21]{GGPLOT2lossSigma_Selected.pdf}} \\
\subfloat[Series]{\includegraphics[page=1,scale=0.21]{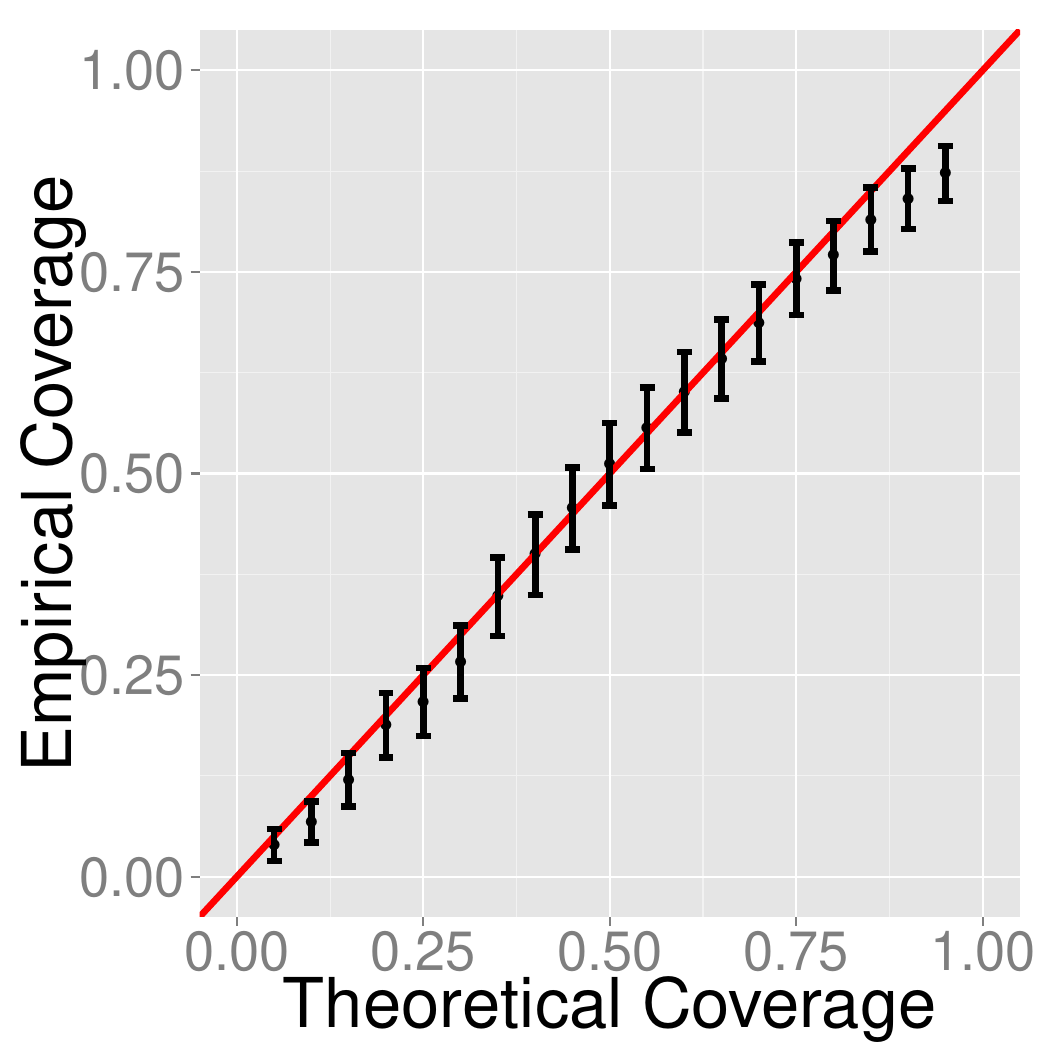}}
\subfloat[NN]{\includegraphics[page=2,scale=0.21]{deep2Coverage.pdf}}
\subfloat[NN-7]{\includegraphics[page=3,scale=0.21]{deep2Coverage.pdf}} \\
\subfloat[Series]{\includegraphics[page=1,scale=0.31]{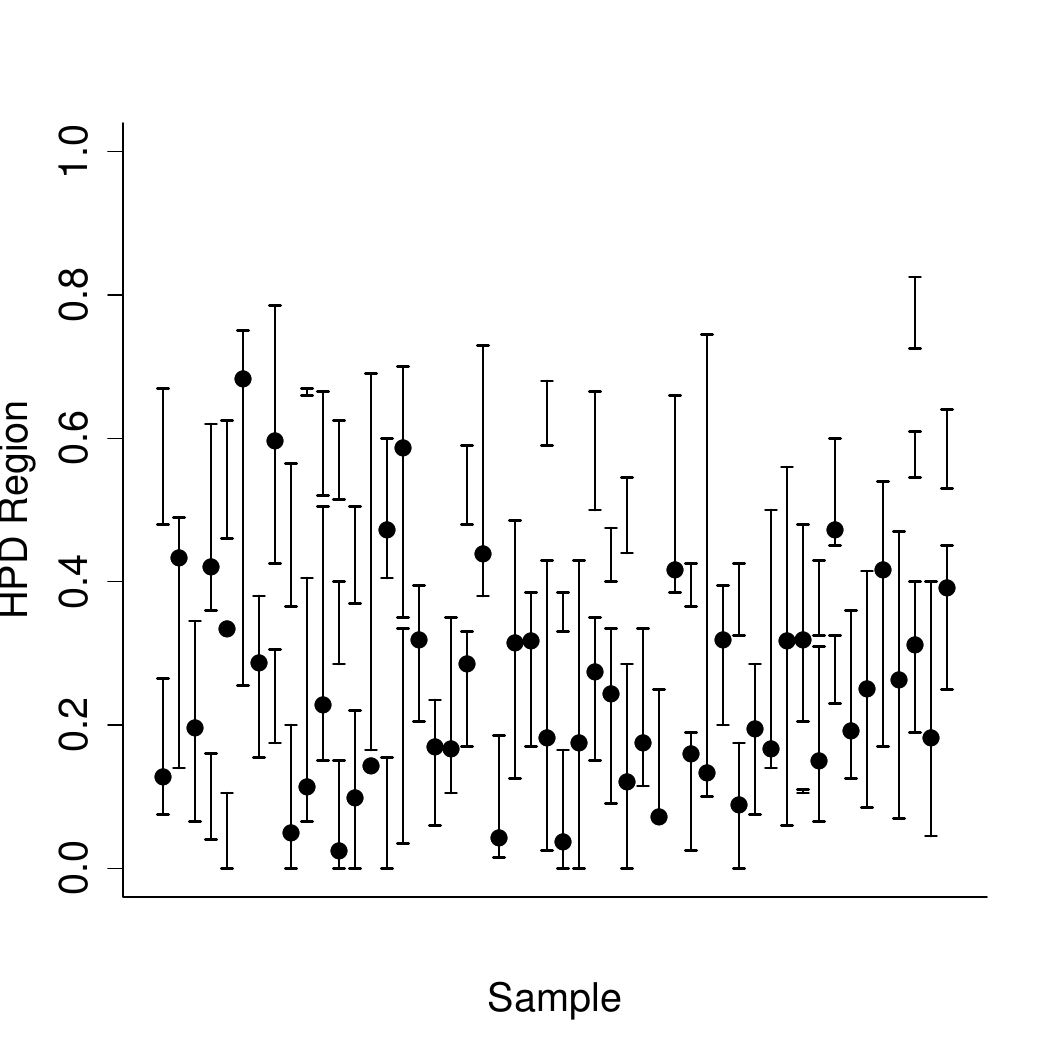}}
\subfloat[NN]{\includegraphics[page=2,scale=0.31]{deep2CoverageHPD.pdf}}
  \caption{Lensing photo-$z$ calibration for DEEP2: \emph{Series} 
  returns parameter estimates with smaller biases and variances than the other approaches (see plots a and b).
 Using 7 neighbors for nearest neighbors as in \citet{Sheldon} yields similar performance in terms of bias; however, the value 7 was hand-picked for this specific task
  and results in poor density estimates (plot e). \emph{Series}  also yields more informative predictive intervals for redshifts than \emph{NN} (plots f and g).}
    \label{fig::estimatesL}
\end{figure}

 Fig.~\ref{fig::estimatesL} shows the results. The nearest neighbor density estimator with 7 neighbors, \emph{NN-7}, has a smaller calibration bias than \emph{NN} with 27 neighbors, the value chosen via the technique described in the paper (see plot a). However, 7 neighbors leads to poor density estimates (see plot e).   \emph{Series} is the only estimator that returns  both  {\em accurate parameter estimates}  for different values of $z_l$ and lenses $l$ (plots a and b), as well as {\em accurate photo-$z$ density estimates} (plots c and f).
  The other estimators do not have both of these properties simultaneously: depending on the chosen tuning parameter, either they have small bias but bad coverage (as \emph{NN-7}), or good coverage but high bias (as \emph{NN}) (see Appendix~\ref{app:diag} for details on how the coverage is computed.).  

With conditional density estimates, we can also construct {\em predictive density regions} for the unknown redshift. The bottom panel in Fig.~\ref{fig::estimatesL} shows the 95\% Highest Predictive Density regions (HPD; see Appendix~\ref{app:diag} for a definition) of the redshift for
\emph{Series} and \emph{NN}. The former HPD region is typically more informative; the size of the HPD's from \emph{Series} are on average $\approx 75\%$ the size of those from \emph{NN}. 
 
 Finally, we note that estimating 
$f(z|\vec{x})$ by NN and then computing $\int g(z)\widehat{f}(z|\vec{x})dz$ is essentially equivalent to
estimating $\E[g(Z)|\vec{x}]$ directly by performing nearest neighbors \emph{regression} of $g(Z)$ on $\vec{x}$. In other words: for this particular application, \emph{NN-7} yields a good estimate of the regression of $g(Z)$ on $\x$, but not (as previously assumed) a reasonable estimate of photo-$z$.  


\comment{ Finally, we note that the regression-based estimator, \emph{Photo-z}, has a large bias, and that
\emph{ker-NN} has a bias in-between \emph{NN} and \emph{Series}. The latter result is in agreement
with our previous experiments: when there is no selection bias, \emph{Series} has better performance
than \emph{ker-NN}.}

\section{Conclusions}
\label{sec-conclusions}

\comment{In this work, we proposed and analyzed new non-parametric methods for estimating conditional densities under selection bias. Specifically, we worked with the Missing
at Random assumption in the context of photometric redshift prediction. 

To match training (spectroscopic) and target (photometric) samples, we used a reweighting scheme based on importance weights. We found that the nearest neighbors estimator from \citet{Cunha1} is 
very effective for estimating these weights,  even when compared to 
state-of-the-art approaches for density ratio estimation from the machine learning literature. We provided a principled way of choosing 
 tuning parameters, 
 and we introduced two new conditional density estimators, \emph{kernel nearest neighbors} (Section \ref{sec-ker-NN}) and \emph{Series} (Section \ref{sec-CDEunderCS}), that both have better performance than the photo-z prediction method by \citet{Cunha1}. }

Over the past 15 years, the number of applications for estimating photometric redshifts have grown rapidly, and today there exist a large number of techniques (or codes) for estimating redshifts; see, e.g., \citet{dahlen2013} and references therein. 
With next-generation surveys, we also expect to have access to additional data (e.g., surface brightness or sizes of galaxies, \citealt{Lima1}, or other magnitudes such as grizYJHKs, \citealt{Oyaizu}) which could potentially improve  current photo-$z$ estimates. The value of our work is that it provides a principled framework for properly tuning and assessing different estimators, as well as  methods for selecting covariates and for combining two or more estimators for optimal performance. 
In this paper, we also compared some new and existing estimators of importance weights and photometric redshift under different settings. We found that the nearest neighbors estimator from \citet{Cunha1} is very effective for estimating importance weights,  even when compared to 
state-of-the-art approaches for density ratio estimation from the machine learning literature.  We introduced two new non-parametric conditional density estimators, \emph{kernel nearest neighbors} (Section \ref{sec-ker-NN}) and \emph{Series} (Section \ref{sec-CDEunderCS}), that both have better performance than the photo-z prediction method by \citet{Cunha1}.  

Finally, although the scope of this paper is conditional density estimation, 
one can directly apply the proposed methods to the regression of functions of the photometric redshift,  as the regression $\E[g(Z)|\vec{x}]=\int g(z) f(z|\vec{x}) d\vec{x}$. The galaxy-galaxy lensing example in Sec.~\ref{sec-lensing} clearly illustrates the importance of properly tuning and assessing photo-$z$ density estimators in down-stream cosmological analysis.  In the example, our estimator yielded more accurate predictive regions for the redshift $z$, as well as better estimates of $g(z)$ for a range of different functions $g$, although we did not explicitly take $g$ into account in the optimization. We believe our proposed techniques will be valuable for astronomers and cosmologists carrying
out next-generation surveys, where accurate photo-$z$ estimates will be
needed within a range of different applications (and functions $g$). 
\vspace{2mm}

{\small \noindent \textbf{Acknowledgments.}
	We thank Jeffrey A. Newman for his insightful comments. 
 We are also grateful to the referees and editors for the detailed comments that helped improve the paper.
 This work was partially supported by \emph{Conselho Nacional de Desenvolvimento Cient\'ifico e Tecnol\'ogico} (200959/2010-7),
 \emph{Funda\c{c}\~ao de Amparo \`a Pesquisa do Estado de S\~ao Paulo} (2014/25302-2), the \emph{Estella Loomis McCandless Professorship}, and NSF DMS-1520786.}\\

\begin{center}
{\small \noindent \textbf{SUPPLEMENTARY MATERIAL}}
\end{center}

We provide the data and code used in the paper as supplementary material.

\bibliographystyle{plainnat}
\bibliography{main}

\newpage

\appendix

\section{Diagnostic Tests for Conditional Density Estimation}
\label{app:diag}

The $L^2$ loss only 
  conveys limited information on how well the final density estimates 
  actually fit the observed data. 
 Below we describe three diagnostic so-called {\em goodness-of-fit} tests that one can use to more closely assess the quality of different models; similar tests can be found in the time series literature (see, e.g., \citealt{Corradi}). Let $\widehat{F}_{z|\vec{x}_i}$ denote the estimated conditional cumulative distribution function for $z$ given $\vec{x}_i$. Then,

\begin{enumerate}
 \item(\textbf{Q-Q Plot}) For every $c$ in a grid of values on $[0,1]$ and for every observation $i$ in the test spectroscopic sample, compute $Q_i^c = \widehat{F}_{z|\vec{x}_i}^{-1}(c)$. Define
 $\widehat{c}=\frac{1}{n}\sum_{i=1}^n\widehat{\beta}(x_i^L)\I(z^L_i \leq Q_i^c).$ We plot the values of $\widehat{c}$ against the corresponding values of $c$. If the distributions $\widehat{F}_{z|\vec{x}}$ and $F_{z|\vec{x}}$ are similar, then the points in the Q-Q plot will approximately lie on the line $\widehat{c} = c$.

 \item(\textbf{P-value})  For every test data point $i$, let 
 $U_i=\widehat{F}_{z|\vec{x}_i}(Z_i).$
 If the data are really distributed according to $\widehat{F}_{z|\vec{x}}$, then 
 $U_1,\ldots,U_n \overset{\mbox{\tiny{iid}}}{\sim} Unif(0,1)$. Hence, we compute the p-value for a Kolmogorov-Smirnoff test that compares the distributions of these statistics to the uniform distribution. 
 
 \item(\textbf{Coverage Plot and HPD Regions}) For every $\alpha$ in a grid of values in  $[0,1]$  and for every spectroscopic data point $i$ in the test sample, let $A_i$ be a set such 
 that 
 $\int_{A_i}\widehat{f}(z|\vec{x}^L_i)dz=\alpha.$ Here we choose the set $A_i$ with the smallest area: $A_i=\{z:f(z|\x_i^L)>t\}$ 
 where $t$ is such that  $\int_{A_i}\widehat{f}(z|\vec{x}^L_i)dz=\alpha$
 (i.e., $A_i$
 is a Highest Predictive Density region; HPD).
 Define 
 $\widehat{\alpha}_i=\frac{1}{n}\sum_{i=1}^n \widehat{\beta}(x_i^L) \I(z^L_i \in A_i).$ 
 If the distributions $\widehat{F}_{z|\vec{x}}$ and $F_{z|\vec{x}}$ are similar, then $\widehat{\alpha} \approx \alpha$.
 Hence, we plot a graph of $\widehat{\alpha}$'s versus $\alpha$'s and assess how close they are to the line $\widehat{\alpha} = \alpha$. For each $\alpha$, we also include a $95\%$ confidence interval
 based on a normal approximation to the binomial distribution.
\end{enumerate}

\section{Estimating Importance Weights}
\label{app:bias}

Some common approaches to estimating density ratios include direct basis 
expansions of the form 
$\widehat{\beta}(\vec{x})=\sum_{i=1}^I \widehat{\alpha_i} \lambda_i(\vec{x})$ 
\citep{Kanamori,SugiyamaImportance,IzbickiLeeSchafer}, 
kernel mean matching (KMM; \citealt{Gretton}),
and various machine learning techniques (see, e.g.,
\citealt{BickelBrucker} and \citealt{Margolis} for a review). Astronomers have also
themselves explored nearest neighbor-based techniques for reweighting 
non-representative training samples. In particular, \citet{Lima1} and \citet{Cunha1} have had success in photo-z estimation with  
the nearest-neighbor estimator
\begin{align}
\label{eq-nnWeights}
\widehat{\beta}(\vec{x})=\frac{1}{M}\frac{n_L}{n_U}\sum_{k=1}^{n_U}\I\left(\x^U_k \in V_\vec{x}^M\right),
\end{align}
where \begin{align*}
V_\vec{x}^M=\{\vec{y} \in \mathbb{R}^d: d(\vec{y},\vec{x})\leq d(\vec{x}^L_{(M)},\vec{x})\}
\end{align*}
denotes the region of feature space with points that are closer to $\vec{x}$ than
$\vec{x}^L_{(M)}$, the $M$th nearest neighbor of $\vec{x}$ among labeled data, is to $\x$.
This estimator has also been used in the machine learning literature
\citep{Loog}, although only for the case $M = 1$.

{ \em Model Selection and Tuning of Parameters.} To select the best method for estimating  $\beta(\x)$, we need to specify an appropriate loss function. Our ultimate goal is good photo-$z$ prediction for new {\em unlabeled} data. If we use importance weighting according to Eq.~\ref{eq:importance_weighting}, then we need good estimates of $\beta(\vec{x})$ {\em at the 
labeled points}, or more generally, in regions where 
the density of labeled points is large. Hence, we define 
the loss function according to Eq.~\ref{eq-lossBeta}, which is weighted with respect to $\P_L$. (These weights are later used in Eq.~\ref{eq-lossCDEEstimate} to estimate the $\P_U$-weighted loss in Eq.~\ref{eq-lossCDE} for CDE.)


\end{document}